\documentclass[]{aa}

\usepackage[utf8]{inputenc}
\usepackage{natbib}
\usepackage{graphicx}
\usepackage{subcaption}
\usepackage{numprint}
\usepackage{color}
\usepackage{txfonts}
\usepackage{hyperref}
\usepackage{tabularx}
\usepackage{amsmath}
\usepackage{amssymb}
\usepackage{siunitx}
\usepackage{mathtools}
\usepackage[normalem]{ulem}
\usepackage{tikz}
\usetikzlibrary{calc,shapes,arrows}
\usepackage{multirow}
\usepackage{threeparttable}
\providecommand{\Unit}[1]{\ensuremath{\mathrm{~#1}}} 

\providecommand{\parallax}{\ensuremath{\varpi}}

\newcommand\secref[1]{Sect.~\ref{#1}}
\newcommand\secrefalt[1]{Section~\ref{#1}}
\newcommand\figref[1]{Fig.~\ref{#1}}

\newcommand\figrefalt[1]{Figure~\ref{#1}}

\newcommand\tabref[1]{Table~\ref{#1}}
\begin{document}

\title{Hunting for open clusters in \textit{Gaia} DR2: the Galactic anticentre\thanks{Table 2 is only available at the CDS}}

\author{
                A. Castro-Ginard                                        \inst{\ref{inst:UB}}\relax
\and    C. Jordi                                        \inst{\ref{inst:UB}}\relax
\and    X. Luri                                 \inst{\ref{inst:UB}}\relax
\and    T. Cantat-Gaudin                                        \inst{\ref{inst:UB}}\relax
\and    L. Balaguer-N\'{u}\~{n}ez                                       \inst{\ref{inst:UB}}\relax
}

\institute{Dept. F\'isica Qu\`antica i Astrof\'isica, Institut de Ci\`encies del Cosmos (ICCUB), Universitat de Barcelona (IEEC-UB), Mart\'i i Franqu\`es 1, E08028 Barcelona, Spain\\
    \email{acastro@fqa.ub.edu}\relax \label{inst:UB}
}

\date{Received date /
Accepted date}

\abstract{
The \textit{Gaia} Data Release 2 (DR2) provided an unprecedented volume of precise astrometric and excellent photometric data. In terms of data mining the \textit{Gaia} catalogue, machine learning methods have shown to be a powerful tool, for instance in the search for   unknown stellar structures. Particularly, supervised and unsupervised learning methods combined together  significantly improves the detection rate of open clusters.
}{
We systematically scan \textit{Gaia} DR2 in a region covering the Galactic anticentre and the Perseus arm $(\ang{120} \leq l \leq \ang{205}$ and $\ang{-10} \leq b \leq \ang{10})$, with the goal of finding any open clusters that may exist in this region, and fine tuning a previously proposed methodology successfully applied to TGAS data, adapting it to different density regions.
}{
Our methodology uses an unsupervised, density-based, clustering algorithm, DBSCAN, that identifies overdensities in the five-dimensional astrometric parameter space $(l,b,\varpi,\mu_{\alpha^*},\mu_{\delta})$ that may correspond to physical clusters. The overdensities are separated into physical clusters (open clusters) or random statistical clusters using an artificial neural network to recognise the isochrone pattern that open clusters show in a colour magnitude diagram.
}{
The method is able to recover more than $75\%$ of the open clusters confirmed in the search area. Moreover, we detected $53$ open clusters unknown previous to \textit{Gaia} DR2, which represents an increase of more than $22\%$ with respect to the already catalogued clusters in this region.
}{
We find that the census of nearby open clusters is not complete. Different machine learning  methodologies for a blind search of open clusters are complementary to each other;  no single method is able to detect  $100\%$ of the existing groups. Our methodology has shown to be a reliable tool for the automatic detection of open clusters, designed to be applied to the full \textit{Gaia} DR2 catalogue. 
}
\maketitle


\section{Introduction}
\label{sec:intro}

The popularity of machine learning (ML) techniques used to analyse astronomical data is growing, as is  the volume of astronomical catalogues. The use of these techniques is mandatory to extract meaningful insight from big data sets  such as the second data release of the  ESA \textit{Gaia} astrometric mission \citep[\textit{Gaia} DR2,][]{2016A&A...595A...1G,2018A&A...616A...1G}, which contains  more than $550$ GB\footnote{https://www.cosmos.esa.int/web/gaia/dr2} of data, including precise astrometry \citep{2018A&A...616A...2L} and excellent photometry \citep{2018A&A...616A...4E}, among other products, for more than $1.3 \times 10^9$ sources down to magnitude $G = 21$ mag. This unprecedented volume of extremely precise data reveals unseen details in the structure of our galaxy. 

Open clusters (OCs) are considered as fundamental objects in our understanding of the structure and evolution of the Milky Way disc. The stars of an OC were born and move together; \textit{i.e.} in terms of \textit{Gaia} observables, they share  $(l,b,\varpi,\mu_{\alpha^*},\mu_{\delta})$ and follow a specific pattern in a colour-magnitude diagram (CMD) $(G,G_{BP},G_{RP})$. That they can represent overdensities in  five-dimensional astrometric space can be exploited by unsupervised learning algorithms to either characterise known OCs when looking for new member stars \citep{2018ApJ...869....9G,2018Ap&SS.363..232G,tristan_catalogue}, or to detect new overdensities in the parameter space \citep{acastro1,coin_clusters}. Supervised learning methods can help in determining whether  a group of stars is an OC by identifying the isochrone pattern of its member stars in a CMD, due to the common age of its members. In the OC domain, \textit{Gaia} DR2 represents a perfect scenario for the application of ML methods to both its detection and characterisation.

Our understanding of the OC population has dramatically changed with \textit{Gaia} DR2. A pre-\textit{Gaia} census of the OC population counted  around $3000$ objects \citep{dias,kharchenko,2007MNRAS.374..399F,2014A&A...568A..51S,2015A&A...581A..39S,2016A&A...595A..22R} compiled from heterogeneous data sources, making the characterisation of OC parameters a difficult task. After the publication of \textit{Gaia} DR2, \citet{tristan_catalogue} revisited the OC population using a ML based unsupervised membership determination algorithm. This resulted in the compilation of a homogeneous OC catalogue of $1229$ objects, including some serendipitously detected OCs and discarding some objects listed in previous catalogues. These well-determined members and mean astrometrical parameters from the \textit{Gaia} DR2 data allowed the kinematical study of these objects \citep{2018A&A...619A.155S} and the derivation of ages and physical parameters \citep{2019A&A...623A.108B}. Additionally, the combination of ML techniques and \textit{Gaia} DR2 data triggered the detection of new OCs. The discovery of nearby OCs \citep{acastro1,coin_clusters}, where the census was thought to be complete, showed the necessity to keep exploring the sky for new objects.

In \citet{acastro1} (hereafter CG18) we presented a method for the automatic detection of OCs in the \textit{Gaia} data. The method consists in the application of an unsupervised clustering algorithm, DBSCAN, that looks for  overdensities in the astrometric five-dimensional space $(l,b,\varpi,\mu_{\alpha^*},\mu_{\delta})$. Once the overdensities are detected, we classify them as either random statistical overdensities or real OCs by identifying the isochrone pattern of OC member stars in a CMD using an artificial neural network (ANN). The method has proved to be successful in the detection of OCs in the TGAS data \citep{gdr1-tgas,tgas}, which were later validated in the \textit{Gaia} DR2 data. In this paper we apply the methodology to a region of the sky around the Galactic anticentre with the aim of increasing our knowledge of the OC population in that region, and fine tuning the methodology for its planned future application in an all sky blind search.

The paper is organised as follows. \secrefalt{sec:method} briefly describes the methodology used, which is discussed in detail in CG18. The data set used for the detection is described in \secref{sec:data}. The proposal of new OCs and some comments on the results found are in \secref{sec:results}. Finally, concluding remarks are summarised in \secref{sec:conclusions}.

\section{Methodology}
\label{sec:method}
This section briefly describes the methodology used in CG18, where our approach to detect OCs in the \textit{Gaia} DR2 data is explained in detail. The method consists of three parts: a preprocessing step, where the data is prepared to be exploited;  a density-based clustering algorithm, DBSCAN \citep{dbscan}, used to look for overdensities in the five-dimensional astrometric data; and a classification of the resulting clusters into real OCs and random statistical clusters using an ANN \citep{ann} to recognise the isochrone pattern of the cluster member stars in a CMD. 

In the preprocessing step the sky area of study is divided into smaller regions,  rectangles of size $L \times L$ deg, in order  to compute a representative average star density of the region used to search for overdensities. In each rectangle the parameters used to perform the clustering analysis ($l,b,\varpi,\mu_{\alpha^*},\mu_{\delta}$), are standardised (re-scaled to have zero mean and variance of one) to avoid a preferred dimension and to balance the importance of each dimension on the clustering process.

The detection of statistical clusters is done using the \mbox{DBSCAN}\footnote{\label{note1}Algorithm from the scikit-learn python package \citep{sklearn}} algorithm, which is a density-based algorithm that uses the notion of distance between stars to define close stars as a cluster. The statistical distance between two stars is computed as the Euclidean distance in the standardised five-dimensional  parameter space. The reasons for the choice of DBSCAN are twofold. Firstly, it is able to detect arbitrarily shaped clusters, so it accounts, for instance, for the effects of the projection of a cluster location into a two-dimensional sky $(l$ and $b)$. Secondly, it requires only two input parameters: $minPts$, the minimum number of stars needed to be considered a cluster,  and $\epsilon$, the radius of the hyper-sphere where we search for these $minPts$ stars. The parameter $\epsilon$ is automatically computed in each rectangle, assuming that the distance between neighbours in a cluster is smaller than that between field stars (see Sect. 2.2 of CG18). 

\begin{figure}
\begin{subfigure}{1.00\columnwidth}
\centering
\includegraphics[scale = .5]{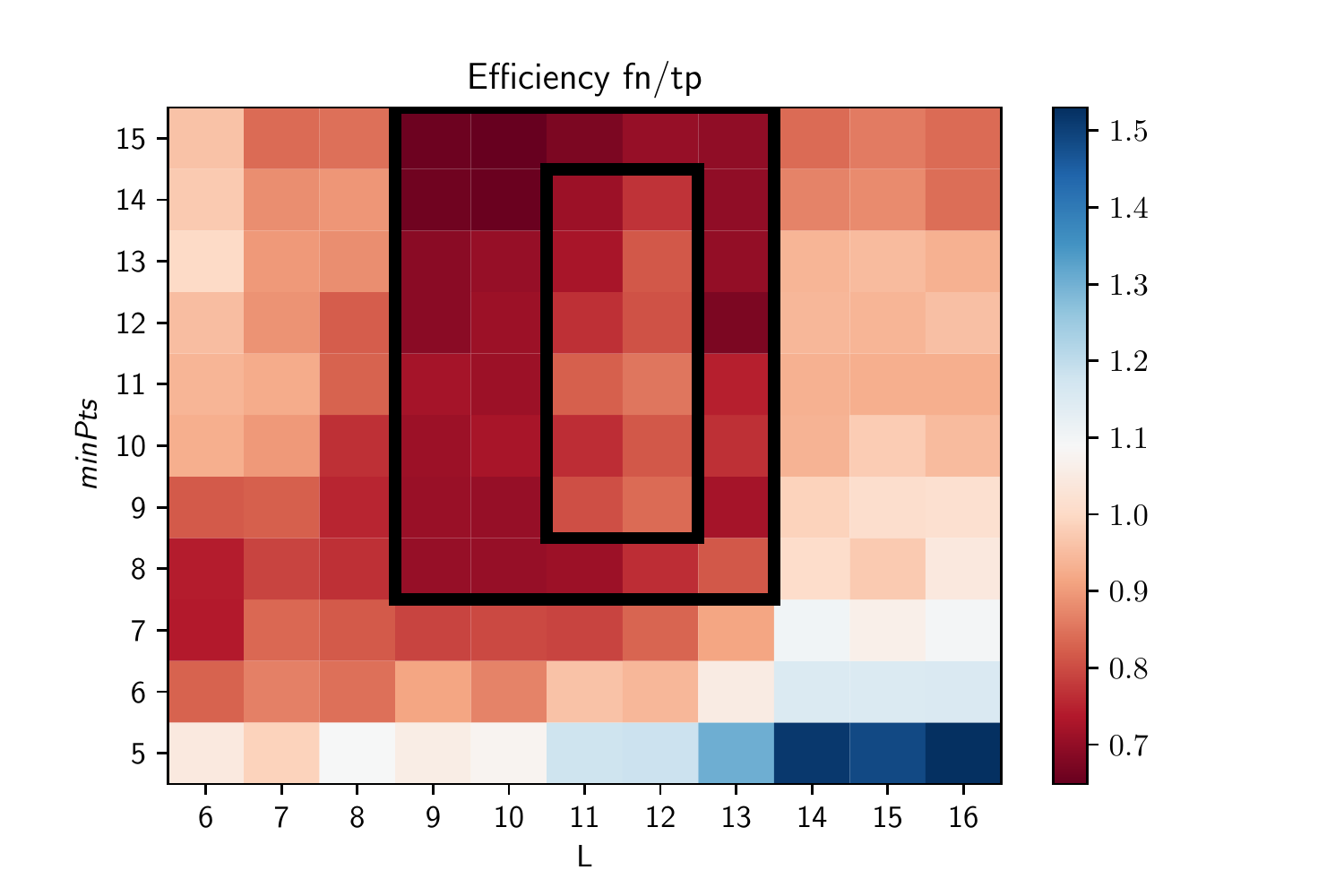}
\end{subfigure}
\begin{subfigure}{1.00\columnwidth}
\centering
\includegraphics[scale = .5]{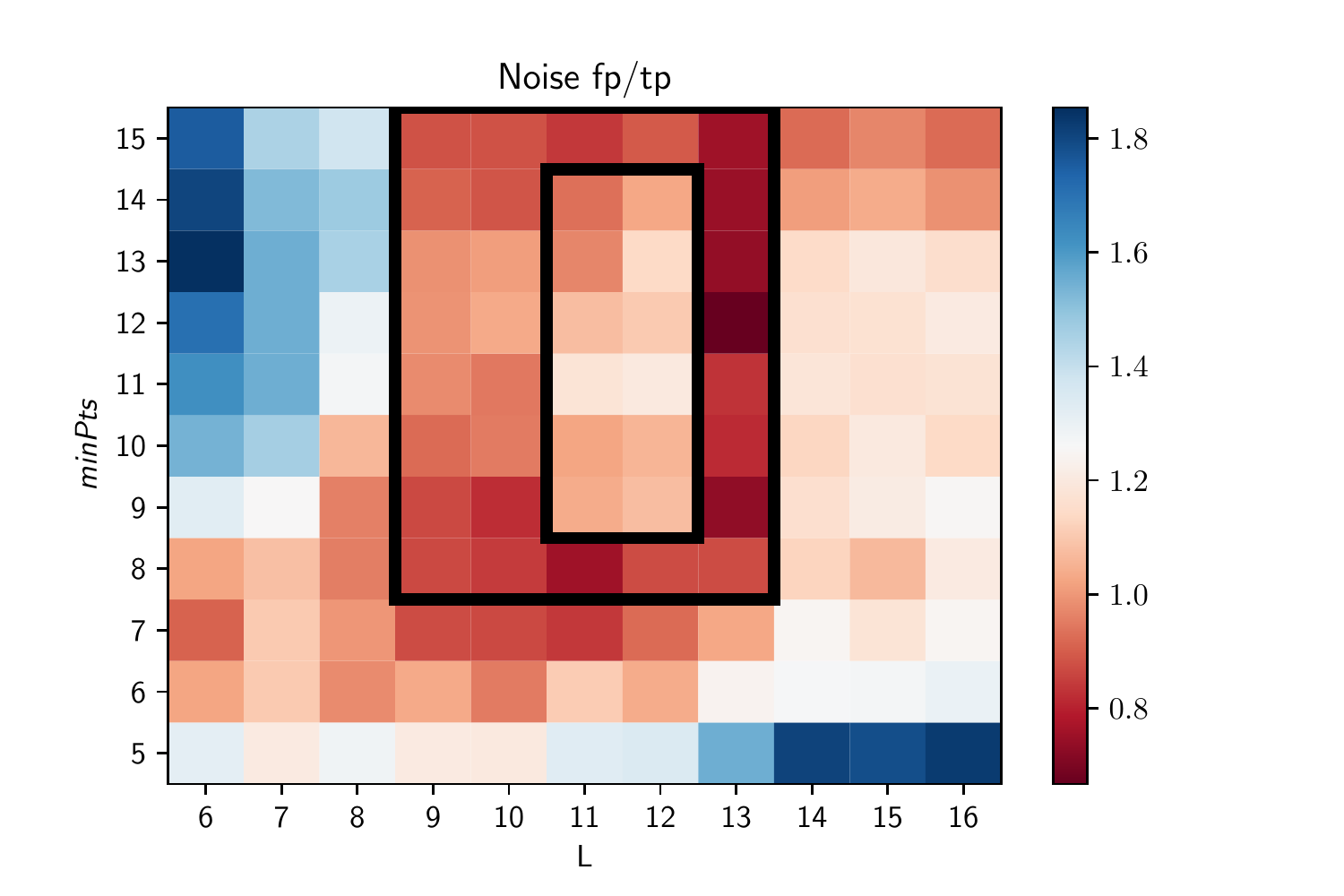}
\end{subfigure}
\caption{Pairs of parameters $(L,minPts)$ explored. The top plot shows the efficiency  of each pair (false negative - true positive rate), whilst the bottom plot shows the noise (false positive - true positive rate). The redder the pixel, the better the performance of the algorithm in terms of OC detection. The parameters selected, considered  optimal for OC detection, are inside the black lines.}
\label{fig:optimalparameters}
\end{figure}

The values for the parameters $(L,minPts)$ are set using \textit{Gaia} DR2-like simulated data (see Sect. 3 in CG18), where in this case we added the errors\footnote{Implementation provided by PyGaia package: https://github.com/agabrown/PyGaia} at the time of \textit{Gaia} DR2. Several combinations of optimal parameters $(L,minPts)$ were selected in order to assess  the resulting performance of the algorithm; in this case we chose $28$ pairs of $(L,minPts)$. \figrefalt{fig:optimalparameters} shows the pairs of parameters explored and the chosen combination inside the black lines, whose values range within $L \in [\ang{9},\ang{13}]$ and $minPts \in [8,15]$. These parameters were selected to try to find a balance between low noise and good efficiency, defined as the false positive - true positive ratio for the noise and false negative - true positive for the efficiency.

After the clustering process, the resulting clusters can be either real OCs or random statistical clusters. These two types can be differentiated by the pattern followed by the cluster member stars on a CMD. The classification into real OCs or random statistical clusters is done with an ANN\textsuperscript{\ref{note1}} that is able to identify the characteristic shape of isochrones in CMDs corresponding to real OCs. To train the ANN we used CMDs from OCs from the most homogeneous OC catalogue to date \citep[see details in][]{tristan_catalogue}, which also has the advantage of being compiled from \textit{Gaia} DR2 data so it is representative of the OCs we expect to detect, and with similar photometric errors. The training set consists of a sample of $1229$ real OCs. In addition we used data augmentation techniques so the volume of the training set was increased by randomly selecting member stars to create a set of subclusters from each of these catalogued OCs. On the negative identification side, we used CMDs from random field stars on the same field as the $1229$ OCs to avoid location biases.

As a last step, and to ensure the selection only of newly detected OCs, we removed the already catalogued OCs. We improved this step with respect to CG18 thanks to the compilation of the catalogue by \citet{tristan_catalogue}. In this case positional arguments were used in order to match a found OC with a catalogued one. An OC was considered to be already catalogued if the mean parameters $(l,b,\varpi,\mu_{\alpha^*},\mu_{\delta})$ of its members was compatible within $2\sigma$ of the mean parameters of the catalogued OC in \citet{tristan_catalogue}. We did not make a cross-identification with other catalogues such as \citet{dias}, \citet{kharchenko} and \citet{2019AJ....157...12B}, and others, due to the inhomogeneous data sources they are compiled from.

\section{Data}
\label{sec:data}

The \textit{Gaia} catalogue, in its second data release \citep[\textit{Gaia} DR2,][]{2018A&A...616A...1G}, provides precise five-dimensional astrometric data (positions, parallax and proper motions) together with magnitudes in three photometric broad bands $(G,G_{BP}$, and $G_{RP})$ for more than $1.3$ billion sources up to $G=21$ mag. In this work we focus on a region located at the disc ($b \in [\ang{-10},\ang{10}]$) near the Galactic anticentre ($l \in [\ang{120},\ang{205}]$) down to magnitude $G=17$ mag, where we find a total of $8\,715\,057$ sources with mean standard uncertainties of $0.07$ mas for the parallax and $0.1$ mas$\cdot$yr$^{-1}$ for proper motions. 

\begin{figure}
\centering
\includegraphics[width = 1.00\columnwidth]{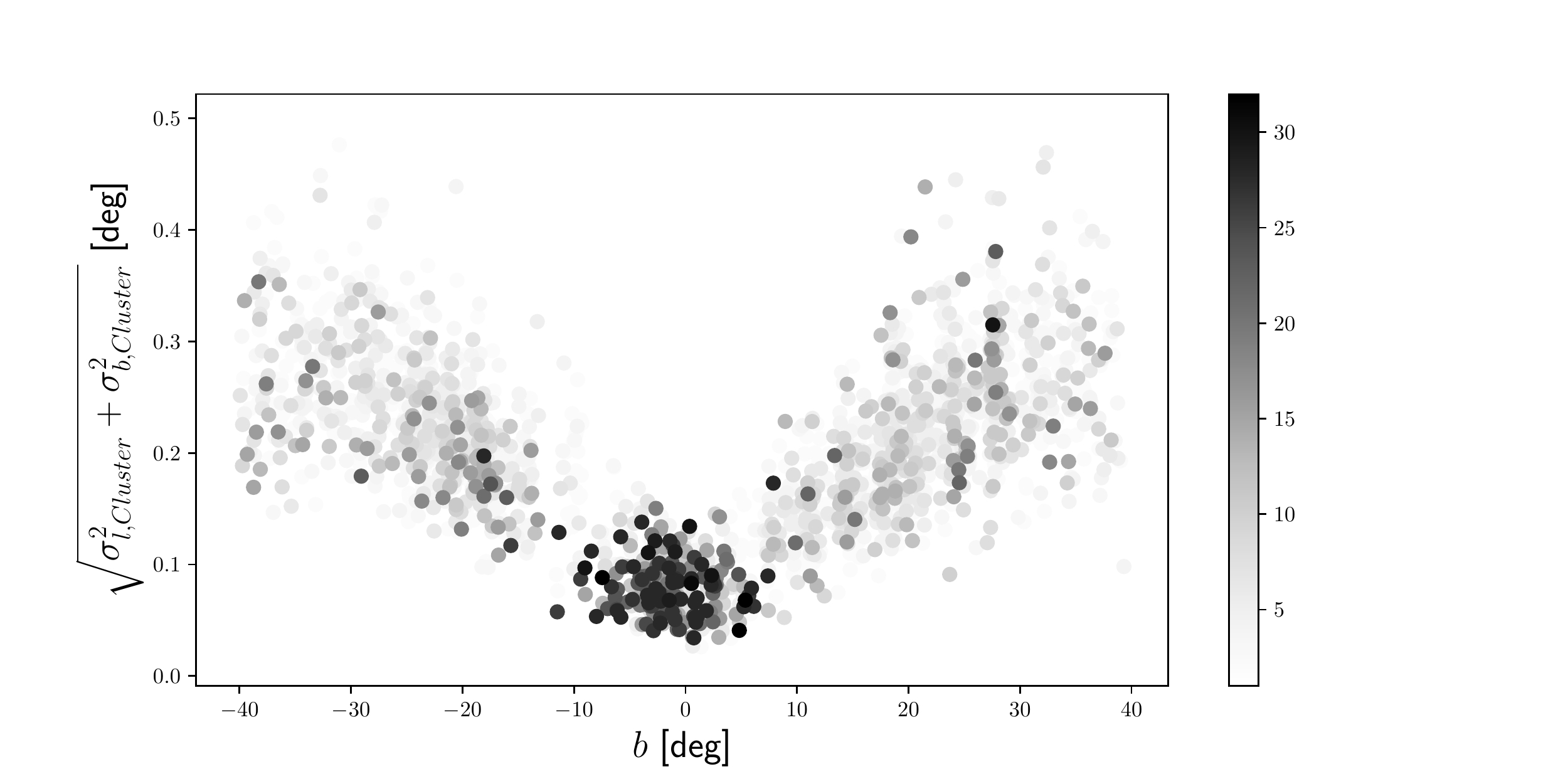}
\caption{Cluster size as a function of the Galactic latitude ($b$). The greyscale represents how many times each cluster is found within the pairs of $(L,minPts)$ explored. High latitude clusters are detected fewer times and are larger in size.}
\label{fig:m10b10just}
\end{figure}

We fixed the search region in the Galactic disc   because the expectation to find OCs decreases at higher altitudes. For instance, around $93\%$ of the OCs catalogued in \citet{tristan_catalogue} are at $|b| < \ang{10}$, and around $99\%$ are located at $|b| < \ang{20}$; similar numbers are found in the catalogues of \citet{dias} and \citet{kharchenko} with $96\%$ and $94\%$ of the OCs located at $|b| < \ang{20}$. Moreover, initially the search region was as wide as $|b| < \ang{40}$, but an exploratory analysis of the results of our method showed that the detection of clusters tends to be less reliable at $|b| > \ang{10}$. This effect is shown in \figref{fig:m10b10just}; the clusters at $|b| > \ang{10}$ are detected fewer  times within the $28$ pairs of $(L,minPts)$ explored than those located at the disc, decreasing the reliability of the candidate. In addition, clusters detected outside the disc increase in size with Galactic latitude, so with decreasing stellar density. Since there is no physical reason for this and although we cannot discard that some of these detected clusters may be real, we interpret that the determination of the $\epsilon$ parameter for such low density regions is not accurate, and therefore we decided to limit the final search region to the disc, defined as $|b| < \ang{10}$.

The reason for the choice of the region near the Galactic anticentre is twofold. On the computational side, the limited volume of data due to the manageable density of stars in the anticentre direction facilitates its analysis, while keeping the richness of the data up to $G = 17$ mag. On the astrophysical side, objects at a greater distance can be reached due to the moderate extinction caused by interstellar dust, compared to the Galactic centre direction. The search region also covers the area recently studied by \citet{coin_clusters} with \textit{Gaia} DR2 data;  they have found $41$ new clusters and note that the region $l \in [\ang{140},\ang{160}]$ seems to be devoid of OCs.

\section{Results}
\label{sec:results}

The method described in \secref{sec:method} is applied to the \textit{Gaia} DR2 data, focused on a region around the anticentre, \textit{i.e.} $\ang{120} \leq l \leq \ang{205}$, and in the disc, $\ang{-10} \leq b \leq \ang{10}$. This results in the detection of $53$ OCs that were unknown previous to \textit{Gaia} DR2, which represent an increase of $\sim 22\%$ with respect to the reference catalogue.

\subsection{Determination of a detection}
\label{subsec:determination}

We can assess the detection criteria by comparing the detected and non-detected OCs from the existing catalogues. In our region of search, \citet{tristan_catalogue} report $240$ OCs of which we were able to recover $182$, \textit{i.e.} $\sim76\%$ of the already known OCs. The reason for the non-detection of the remaining $\sim 24\%$ OCs is related to the contrast of the OC with respect to the field, as seen by the DBSCAN algorithm. 

\begin{figure}
\centering
\includegraphics[width = 1.00\columnwidth]{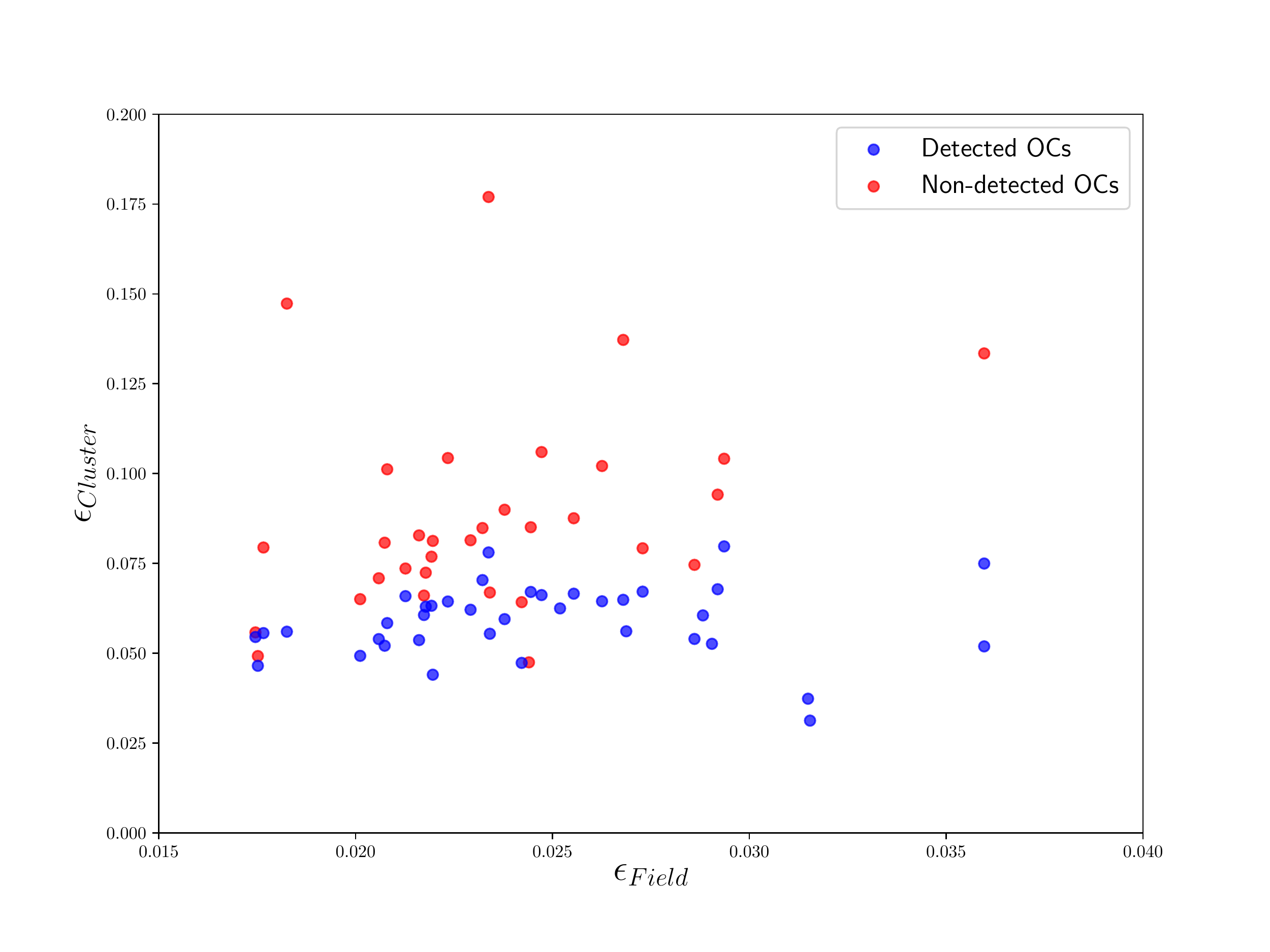}
\caption{Parameter {\large $\epsilon$}  computed for detected (blue) and non-detected (red) OCs in \citet{tristan_catalogue} as a function of the {\large $\epsilon$} computed for the whole field, corresponding to $L = \ang{13}$ and $minPts = 9$.}
\label{fig:determinationdetection}
\end{figure}

\figrefalt{fig:determinationdetection} shows a distribution of the $\epsilon$ parameter computed for each of the $240$ OCs, including  the detected  and non-detected OCs  for $L = \ang{13}$ and $minPts = 9$. The computation of the $\epsilon$ parameter, as explained in Sect. 2.2 of CG18, was done via a data-driven approach; for the interpretation of the parameter the whole data set used has to be taken into account and not just the physical properties of the OCs (or the field). In this case, the key factor that enables the detection of the OC is the OC-field contrast in terms of compactness. We see from \figref{fig:determinationdetection} that only clusters with low values of $\epsilon$ (high contrast) are detected. This is confirmed by the fact that the re-application of the method detects most of the undetected OCs when increasing the contrast with respect to the field by localising the search area to a cone search centred at the targeted OC instead of the large rectangle.

\subsection{Proposal of new OCs}
\label{subsec:newOCs}

The application of our method to the described data set gave us an initial list of $491$ OC candidates. The Monte Carlo-type analysis (application of the method for several optimal pairs of parameters) allowed us to assess the reliability of these detections by the number of times each cluster was found. In order to clean the initial list from false positives, we manually inspected each of the OC candidates and tried to re-detect the candidate in a smaller field (cone search around the centre of the targeted OC) where the OC field contrast is higher. This re-detection was done using  the DBSCAN algorithm again in a cone search region centred on the targeted OC\footnote{For an internal check, we  studied the areas centred on the new OCs with UPMASK \citep{2014A&A...561A..57K} and we confirmed our findings in   $96\%$ of the cases}. The decision on the proposal of the candidate as an OC is made based on the reliability of the candidate and its re-detection. 

After this manual step, $53$ of these $491$ candidates were  validated and proposed as OCs. 
The reason why only $53$ OCs were validated is related to the low complexity of the ANN  architecture, and the low volume of training data available (based on \textit{Gaia} DR2 data only). This can give false positive identifications, \textit{i.e.} incorrect identification,  of a stellar structure as an OC. In our manual validation step we were conservative, tending to accept as OCs only those groups without a sparse distribution in the sky, with  greater compactness in proper motions and parallax, and with  better defined sequences in the CMD. This step may have  introduced a strong bias in the selection and rejected true clusters. With the improved proper motions and parallaxes of \textit{Gaia} DR3 and a more populated training data set, it will be possible to repeat the analysis to fainter magnitudes and   will produce fewer dubious cases. Even though the method is devised to require  minimal user intervention, this is an important step as the exploitation of the \textit{Gaia} data in terms of blind search for stellar structures is at its initial stages, so a robust OC  catalogue needs to be built to reliably train an automatic detection procedure.

A final list of $53$ OCs is proposed, divided into class A  and class B depending on the reliability of the candidate. Positions $(\alpha,\delta)$ and $(l,b)$ together with mean parameters $(\varpi,\mu_{\alpha^*},\mu_{\delta})$ and mean $V_{rad}$ when available can be found in \tabref{tab:new_oc} for each of the new OCs, which also includes the computed apparent size of the OC and its estimated distance with a one-sigma (asymmetric) confidence interval. A list of the detected members for all the reported OCs is available in Table 2\footnote{available online at VizieR service}.

\begin{table*}
\caption{List of the proposed OCs ordered by increasing $l$. The parameters shown are the mean and standard deviation for the ($N$) members found,  the computed apparent size ($\theta$) and estimated distance ($d$) with one-sigma confidence interval; radial velocity is included when available and  is computed with $N_{V_{rad}}$ members. The name follows the numeration started in CG18.}
\label{tab:new_oc}
\centering
\resizebox{\textwidth}{!}{
\renewcommand{\arraystretch}{1.15}
\begin{threeparttable}[b]
\begin{tabular}{cccccccccccc}
\hline
\hline
\multicolumn{1}{c}{Name} &  \multicolumn{1}{c}{\begin{tabular}[c]{@{}c@{}}$\alpha$\\ 
$[\Unit{deg}]$\end{tabular}} &\multicolumn{1}{c}{\begin{tabular}[c]{@{}c@{}}$\delta$\\ 
$[\Unit{deg}]$\end{tabular}} & \multicolumn{1}{c}{\begin{tabular}[c]{@{}c@{}}$l$\\ 
$[\Unit{deg}]$\end{tabular}} & \multicolumn{1}{c}{\begin{tabular}[c]{@{}c@{}}$b$\\ 
$[\Unit{deg}]$\end{tabular}} & \multicolumn{1}{c}{\begin{tabular}[c]{@{}c@{}}$\theta$\\ 
$[\Unit{deg}]$\end{tabular}} & \multicolumn{1}{c}{\begin{tabular}[c]{@{}c@{}}$\parallax$\\ 
$[\Unit{mas}]$\end{tabular}} & \multicolumn{1}{c}{\begin{tabular}[c]{@{}c@{}}$d$\\ 
$[\Unit{kpc}]$\end{tabular}} & \multicolumn{1}{c}{\begin{tabular}[c]{@{}c@{}}$\mu_{\alpha^*}$\\ 
$[\Unit{mas\cdot yr}^{-1}]$\end{tabular}} & \multicolumn{1}{c}{\begin{tabular}[c]{@{}c@{}}$\mu_{\delta}$\\ 
$[\Unit{mas\cdot yr}^{-1}]$\end{tabular}}  & \multicolumn{1}{c}{\begin{tabular}[c]{@{}c@{}}$V_{\rm rad}$\\ 
$[\Unit{km\cdot s}^{-1}]$\end{tabular}}  & \multicolumn{1}{c}{$N$ ($N_{V_{\rm rad}}$)} \\ 
\hline 
\multicolumn{10}{c}{Class A}\\ 
\hline 
\object{UBC~33}&$7.39$$(0.18)$&$60.49$$(0.08)$&$120.24$$(0.09)$&$-2.27$$(0.08)$&$0.12$&$0.63$$(0.03)$&$1.6^{+0.08}_{-0.07}$&$-0.94$$(0.09)$&$-0.46$$(0.06)$&$-$$(-)$&$43$$(0)$\\ 
\object{UBC~34}\tnote{a}&$11.8$$(0.22)$&$66.75$$(0.15)$&$122.51$$(0.09)$&$3.89$$(0.15)$&$0.17$&$1.55$$(0.04)$&$0.64^{+0.02}_{-0.02}$&$-5.02$$(0.31)$&$-3.1$$(0.36)$&$-12.02$$(-)$&$41$$(1)$\\ 
\object{UBC~35}\tnote{a}&$15.1$$(0.11)$&$55.41$$(0.09)$&$124.21$$(0.07)$&$-7.44$$(0.09)$&$0.11$&$0.79$$(0.05)$&$1.27^{+0.08}_{-0.07}$&$-4.46$$(0.2)$&$-1.94$$(0.15)$&$-31.58$$(0.68)$&$70$$(3)$\\ 
\object{UBC~36}&$16.47$$(0.06)$&$59.64$$(0.06)$&$124.76$$(0.03)$&$-3.18$$(0.06)$&$0.07$&$0.47$$(0.04)$&$2.15^{+0.19}_{-0.16}$&$-1.21$$(0.19)$&$-0.46$$(0.14)$&$-50.68$$(5.09)$&$27$$(2)$\\ 
\object{UBC~37}\tnote{a}&$20.95$$(0.38)$&$70.58$$(0.12)$&$125.64$$(0.13)$&$7.88$$(0.12)$&$0.17$&$1.33$$(0.06)$&$0.75^{+0.04}_{-0.03}$&$-6.13$$(0.36)$&$2.08$$(0.27)$&$-25.02$$(1.52)$&$82$$(2)$\\ 
\object{UBC~38}\tnote{a}&$18.73$$(0.11)$&$60.5$$(0.07)$&$125.82$$(0.06)$&$-2.24$$(0.06)$&$0.09$&$0.79$$(0.04)$&$1.27^{+0.06}_{-0.06}$&$-2.45$$(0.13)$&$-1.81$$(0.12)$&$87.09$$(-)$&$56$$(1)$\\ 
\object{UBC~39}&$19.79$$(0.12)$&$61.02$$(0.07)$&$126.29$$(0.06)$&$-1.67$$(0.07)$&$0.09$&$0.48$$(0.03)$&$2.09^{+0.16}_{-0.14}$&$-1.23$$(0.08)$&$-0.13$$(0.12)$&$-$$(-)$&$45$$(0)$\\ 
\object{UBC~40}&$22.63$$(0.06)$&$60.24$$(0.04)$&$127.77$$(0.03)$&$-2.27$$(0.04)$&$0.05$&$0.4$$(0.03)$&$2.48^{+0.19}_{-0.16}$&$-1.01$$(0.24)$&$-0.56$$(0.13)$&$-$$(-)$&$27$$(0)$\\ 
\object{UBC~41}&$23.23$$(0.09)$&$59.79$$(0.05)$&$128.13$$(0.04)$&$-2.66$$(0.05)$&$0.06$&$0.38$$(0.04)$&$2.62^{+0.28}_{-0.23}$&$-0.76$$(0.29)$&$-0.73$$(0.22)$&$-42.02$$(-)$&$47$$(1)$\\ 
\object{UBC~42}\tnote{a}&$26.14$$(0.11)$&$58.74$$(0.05)$&$129.79$$(0.06)$&$-3.42$$(0.05)$&$0.07$&$0.45$$(0.03)$&$2.23^{+0.16}_{-0.14}$&$-0.93$$(0.15)$&$-1.01$$(0.13)$&$-$$(-)$&$55$$(0)$\\ 
\object{UBC~43}\tnote{a}&$28.1$$(0.09)$&$58.65$$(0.06)$&$130.8$$(0.05)$&$-3.29$$(0.06)$&$0.08$&$0.28$$(0.04)$&$3.54^{+0.56}_{-0.43}$&$-2.37$$(0.13)$&$-0.44$$(0.12)$&$-43.7$$(2.34)$&$73$$(2)$\\ 
\object{UBC~44}&$31.11$$(0.1)$&$54.36$$(0.06)$&$133.53$$(0.06)$&$-7.01$$(0.06)$&$0.08$&$0.35$$(0.04)$&$2.84^{+0.32}_{-0.26}$&$-2.2$$(0.24)$&$-0.23$$(0.23)$&$-38.03$$(0.98)$&$47$$(5)$\\ 
\object{UBC~45}\tnote{a}&$33.75$$(0.1)$&$58.45$$(0.04)$&$133.7$$(0.05)$&$-2.67$$(0.04)$&$0.07$&$0.63$$(0.04)$&$1.59^{+0.1}_{-0.09}$&$-1.02$$(0.16)$&$-1.53$$(0.15)$&$-$$(-)$&$31$$(0)$\\ 
\object{UBC~46}&$33.69$$(0.15)$&$57.31$$(0.11)$&$134.03$$(0.08)$&$-3.76$$(0.11)$&$0.14$&$0.4$$(0.03)$&$2.52^{+0.21}_{-0.18}$&$-0.82$$(0.21)$&$-1.14$$(0.22)$&$-$$(-)$&$65$$(0)$\\ 
\object{UBC~47}&$42.0$$(0.09)$&$63.8$$(0.06)$&$135.37$$(0.05)$&$3.78$$(0.05)$&$0.07$&$0.65$$(0.04)$&$1.54^{+0.09}_{-0.08}$&$1.19$$(0.25)$&$-1.12$$(0.17)$&$-10.43$$(-)$&$24$$(1)$\\ 
\object{UBC~48}\tnote{a}&$39.07$$(0.19)$&$50.05$$(0.16)$&$139.64$$(0.14)$&$-9.39$$(0.15)$&$0.2$&$1.36$$(0.05)$&$0.73^{+0.03}_{-0.03}$&$2.5$$(0.31)$&$-2.5$$(0.26)$&$-14.04$$(9.46)$&$49$$(3)$\\ 
\object{UBC~49}&$60.22$$(0.12)$&$59.19$$(0.06)$&$145.14$$(0.07)$&$4.75$$(0.04)$&$0.09$&$0.34$$(0.05)$&$2.97^{+0.56}_{-0.41}$&$-1.77$$(0.13)$&$-1.33$$(0.14)$&$-14.29$$(-)$&$47$$(1)$\\ 
\object{UBC~50}\tnote{a}&$51.5$$(0.13)$&$51.08$$(0.1)$&$146.11$$(0.08)$&$-4.7$$(0.1)$&$0.13$&$0.8$$(0.04)$&$1.25^{+0.07}_{-0.06}$&$2.03$$(0.18)$&$-6.78$$(0.21)$&$-8.78$$(0.3)$&$52$$(2)$\\ 
\object{UBC~51}&$59.67$$(0.17)$&$52.56$$(0.09)$&$149.24$$(0.09)$&$-0.47$$(0.1)$&$0.14$&$0.88$$(0.03)$&$1.14^{+0.04}_{-0.04}$&$-0.23$$(0.23)$&$-1.37$$(0.28)$&$-0.22$$(-)$&$34$$(1)$\\ 
\object{UBC~52}&$64.74$$(0.13)$&$52.37$$(0.11)$&$151.65$$(0.11)$&$1.47$$(0.07)$&$0.13$&$0.41$$(0.04)$&$2.43^{+0.26}_{-0.22}$&$-0.86$$(0.12)$&$0.58$$(0.1)$&$-27.83$$(7.35)$&$32$$(2)$\\ 
\object{UBC~53}&$59.82$$(0.09)$&$47.4$$(0.06)$&$152.68$$(0.06)$&$-4.33$$(0.06)$&$0.08$&$0.6$$(0.04)$&$1.67^{+0.13}_{-0.11}$&$0.67$$(0.12)$&$-2.92$$(0.15)$&$-18.13$$(7.71)$&$47$$(3)$\\ 
\object{UBC~54}&$64.72$$(0.19)$&$46.44$$(0.15)$&$155.8$$(0.14)$&$-2.77$$(0.14)$&$0.2$&$0.88$$(0.05)$&$1.14^{+0.08}_{-0.07}$&$3.33$$(0.23)$&$-3.79$$(0.3)$&$-15.46$$(0.46)$&$143$$(2)$\\ 
\object{UBC~56}&$69.88$$(0.14)$&$47.53$$(0.12)$&$157.43$$(0.12)$&$0.53$$(0.09)$&$0.15$&$1.11$$(0.04)$&$0.9^{+0.03}_{-0.03}$&$1.62$$(0.28)$&$-4.01$$(0.25)$&$-$$(-)$&$72$$(0)$\\ 
\object{UBC~57}&$62.96$$(0.1)$&$42.72$$(0.05)$&$157.48$$(0.06)$&$-6.32$$(0.06)$&$0.09$&$0.48$$(0.05)$&$2.08^{+0.23}_{-0.19}$&$3.19$$(0.22)$&$-2.24$$(0.19)$&$5.24$$(0.23)$&$36$$(3)$\\ 
\object{UBC~58}\tnote{a}&$68.41$$(0.13)$&$40.5$$(0.1)$&$161.94$$(0.11)$&$-4.98$$(0.09)$&$0.14$&$0.95$$(0.06)$&$1.05^{+0.07}_{-0.06}$&$2.03$$(0.41)$&$-3.41$$(0.46)$&$1.0$$(-)$&$39$$(1)$\\ 
\object{UBC~59}&$82.24$$(0.12)$&$48.04$$(0.09)$&$162.06$$(0.1)$&$7.44$$(0.08)$&$0.12$&$0.38$$(0.04)$&$2.62^{+0.35}_{-0.27}$&$0.69$$(0.24)$&$-2.0$$(0.26)$&$-29.73$$(9.06)$&$76$$(5)$\\ 
\object{UBC~60}\tnote{a}&$68.13$$(0.2)$&$39.5$$(0.13)$&$162.54$$(0.16)$&$-5.81$$(0.13)$&$0.2$&$1.47$$(0.05)$&$0.68^{+0.02}_{-0.02}$&$3.62$$(0.43)$&$-5.73$$(0.36)$&$-9.52$$(13.66)$&$71$$(8)$\\ 
\object{UBC~61}&$75.06$$(0.15)$&$36.27$$(0.15)$&$168.55$$(0.15)$&$-3.72$$(0.12)$&$0.19$&$0.75$$(0.05)$&$1.33^{+0.1}_{-0.09}$&$2.1$$(0.14)$&$-2.17$$(0.12)$&$10.1$$(0.93)$&$52$$(2)$\\ 
\object{UBC~62}\tnote{a}&$76.11$$(0.12)$&$35.82$$(0.08)$&$169.42$$(0.09)$&$-3.32$$(0.09)$&$0.13$&$0.83$$(0.05)$&$1.21^{+0.08}_{-0.07}$&$0.36$$(0.18)$&$-3.75$$(0.19)$&$-$$(-)$&$94$$(0)$\\ 
\object{UBC~63}&$79.67$$(0.09)$&$37.82$$(0.08)$&$169.49$$(0.08)$&$0.16$$(0.07)$&$0.11$&$0.65$$(0.04)$&$1.54^{+0.1}_{-0.09}$&$1.12$$(0.18)$&$-3.56$$(0.17)$&$-$$(-)$&$26$$(0)$\\  
\object{UBC~65}\tnote{a}&$82.18$$(0.15)$&$34.32$$(0.14)$&$173.53$$(0.15)$&$-0.15$$(0.1)$&$0.18$&$0.78$$(0.06)$&$1.28^{+0.1}_{-0.09}$&$-1.48$$(0.14)$&$-4.67$$(0.2)$&$-$$(-)$&$79$$(0)$\\ 
\object{UBC~66}\tnote{a}&$78.58$$(0.1)$&$31.72$$(0.09)$&$173.95$$(0.09)$&$-4.1$$(0.09)$&$0.13$&$0.91$$(0.04)$&$1.09^{+0.05}_{-0.04}$&$0.52$$(0.21)$&$-1.48$$(0.23)$&$-$$(-)$&$27$$(0)$\\ 
\object{UBC~67}\tnote{a}&$81.87$$(0.06)$&$33.53$$(0.06)$&$174.05$$(0.05)$&$-0.79$$(0.06)$&$0.08$&$0.47$$(0.03)$&$2.14^{+0.15}_{-0.13}$&$0.45$$(0.13)$&$-2.71$$(0.13)$&$-$$(-)$&$38$$(0)$\\ 
\object{UBC~68}&$91.17$$(0.1)$&$36.77$$(0.07)$&$175.21$$(0.07)$&$7.39$$(0.08)$&$0.1$&$0.43$$(0.05)$&$2.32^{+0.32}_{-0.25}$&$-0.5$$(0.22)$&$-1.69$$(0.23)$&$-$$(-)$&$54$$(0)$\\ 
\object{UBC~69}\tnote{a}&$84.77$$(0.08)$&$28.4$$(0.1)$&$179.7$$(0.1)$&$-1.5$$(0.06)$&$0.12$&$0.71$$(0.04)$&$1.42^{+0.09}_{-0.08}$&$-0.13$$(0.18)$&$-3.82$$(0.22)$&$-$$(-)$&$44$$(0)$\\ 
\object{UBC~70}\tnote{a}&$91.06$$(0.07)$&$31.61$$(0.06)$&$179.72$$(0.06)$&$4.81$$(0.06)$&$0.09$&$0.48$$(0.05)$&$2.07^{+0.23}_{-0.19}$&$-0.72$$(0.16)$&$-3.27$$(0.13)$&$14.57$$(0.79)$&$60$$(2)$\\ 
\object{UBC~72}&$90.99$$(0.09)$&$26.65$$(0.08)$&$184.02$$(0.09)$&$2.34$$(0.08)$&$0.12$&$0.52$$(0.04)$&$1.93^{+0.16}_{-0.14}$&$0.36$$(0.13)$&$-0.01$$(0.15)$&$30.35$$(0.63)$&$77$$(3)$\\ 
\object{UBC~74}&$95.47$$(0.07)$&$22.41$$(0.06)$&$189.7$$(0.06)$&$3.9$$(0.06)$&$0.09$&$0.35$$(0.05)$&$2.82^{+0.45}_{-0.34}$&$1.09$$(0.11)$&$-2.62$$(0.13)$&$43.98$$(1.53)$&$65$$(3)$\\ 
\object{UBC~75}\tnote{a}&$83.77$$(0.07)$&$15.71$$(0.09)$&$190.02$$(0.09)$&$-9.02$$(0.07)$&$0.11$&$0.67$$(0.05)$&$1.5^{+0.11}_{-0.1}$&$0.26$$(0.17)$&$-2.4$$(0.2)$&$5.95$$(-)$&$57$$(1)$\\ 
\object{UBC~76}&$89.0$$(0.11)$&$17.34$$(0.08)$&$191.2$$(0.08)$&$-3.87$$(0.1)$&$0.13$&$0.57$$(0.02)$&$1.75^{+0.07}_{-0.06}$&$0.14$$(0.14)$&$-1.11$$(0.09)$&$-$$(-)$&$24$$(0)$\\ 
\object{UBC~78}\tnote{a}&$85.75$$(0.13)$&$13.72$$(0.1)$&$192.74$$(0.12)$&$-8.41$$(0.1)$&$0.16$&$0.91$$(0.05)$&$1.1^{+0.06}_{-0.05}$&$0.64$$(0.35)$&$-3.65$$(0.33)$&$27.84$$(20.2)$&$62$$(2)$\\ 
\object{UBC~80}&$91.64$$(0.09)$&$8.75$$(0.1)$&$199.97$$(0.1)$&$-5.84$$(0.09)$&$0.14$&$0.45$$(0.04)$&$2.22^{+0.24}_{-0.2}$&$-0.48$$(0.09)$&$-1.01$$(0.21)$&$-$$(-)$&$30$$(0)$\\ 
\object{UBC~81}\tnote{a}&$96.35$$(0.07)$&$11.15$$(0.06)$&$200.05$$(0.06)$&$-0.62$$(0.06)$&$0.09$&$0.58$$(0.03)$&$1.71^{+0.09}_{-0.08}$&$-1.11$$(0.15)$&$-0.94$$(0.14)$&$-$$(-)$&$49$$(0)$\\ 
\object{UBC~82}&$95.89$$(0.08)$&$8.38$$(0.1)$&$202.3$$(0.1)$&$-2.31$$(0.09)$&$0.13$&$0.42$$(0.04)$&$2.38^{+0.28}_{-0.23}$&$1.31$$(0.09)$&$-2.3$$(0.17)$&$12.69$$(0.58)$&$36$$(3)$\\ 
\object{UBC~83}&$97.56$$(0.1)$&$7.36$$(0.12)$&$203.96$$(0.11)$&$-1.33$$(0.12)$&$0.16$&$0.48$$(0.06)$&$2.11^{+0.29}_{-0.23}$&$-1.08$$(0.14)$&$0.39$$(0.05)$&$-$$(-)$&$51$$(0)$\\ 
\hline 
\multicolumn{10}{c}{Class B}\\ 
\hline 
\object{UBC~84}&$15.42$$(0.11)$&$61.73$$(0.09)$&$124.14$$(0.05)$&$-1.11$$(0.09)$&$0.1$&$0.37$$(0.03)$&$2.73^{+0.27}_{-0.23}$&$-1.55$$(0.26)$&$-0.97$$(0.18)$&$-$$(-)$&$55$$(0)$\\ 
\object{UBC~85}&$18.68$$(0.18)$&$57.86$$(0.11)$&$126.04$$(0.09)$&$-4.87$$(0.11)$&$0.15$&$0.36$$(0.03)$&$2.78^{+0.29}_{-0.24}$&$-3.69$$(0.15)$&$-0.54$$(0.33)$&$-$$(-)$&$33$$(0)$\\ 
\object{UBC~86}&$33.03$$(0.16)$&$57.61$$(0.07)$&$133.6$$(0.1)$&$-3.58$$(0.06)$&$0.11$&$0.34$$(0.04)$&$2.93^{+0.4}_{-0.31}$&$-0.85$$(0.16)$&$-0.95$$(0.15)$&$-39.91$$(17.43)$&$71$$(4)$\\ 
\object{UBC~87}&$60.51$$(0.1)$&$56.42$$(0.07)$&$147.09$$(0.07)$&$2.77$$(0.06)$&$0.09$&$0.38$$(0.03)$&$2.62^{+0.25}_{-0.21}$&$0.77$$(0.15)$&$-1.31$$(0.16)$&$-$$(-)$&$36$$(0)$\\ 
\object{UBC~88}&$58.18$$(0.18)$&$45.94$$(0.15)$&$152.76$$(0.14)$&$-6.17$$(0.14)$&$0.2$&$1.0$$(0.06)$&$1.0^{+0.06}_{-0.05}$&$-1.36$$(0.33)$&$-2.95$$(0.27)$&$-$$(-)$&$88$$(0)$\\ 
\object{UBC~89}\tnote{a}&$81.22$$(0.14)$&$37.57$$(0.08)$&$170.4$$(0.09)$&$1.03$$(0.1)$&$0.14$&$0.88$$(0.06)$&$1.13^{+0.08}_{-0.07}$&$0.39$$(0.23)$&$-4.27$$(0.22)$&$59.54$$(-)$&$64$$(1)$\\ 
\object{UBC~90}&$97.21$$(0.04)$&$14.92$$(0.05)$&$197.11$$(0.05)$&$1.87$$(0.04)$&$0.06$&$0.34$$(0.05)$&$2.96^{+0.5}_{-0.37}$&$1.23$$(0.14)$&$-1.38$$(0.16)$&$49.63$$(-)$&$53$$(1)$\\ 
\hline 
\end{tabular}\begin{tablenotes}
\item[a] coincidence with COIN clusters
\end{tablenotes}
\end{threeparttable}
}
\end{table*} 

\subsection{Comments on the detected OCs}
\label{subsec:comments}

\begin{figure*}
\centering
\includegraphics[width = 1.00\textwidth]{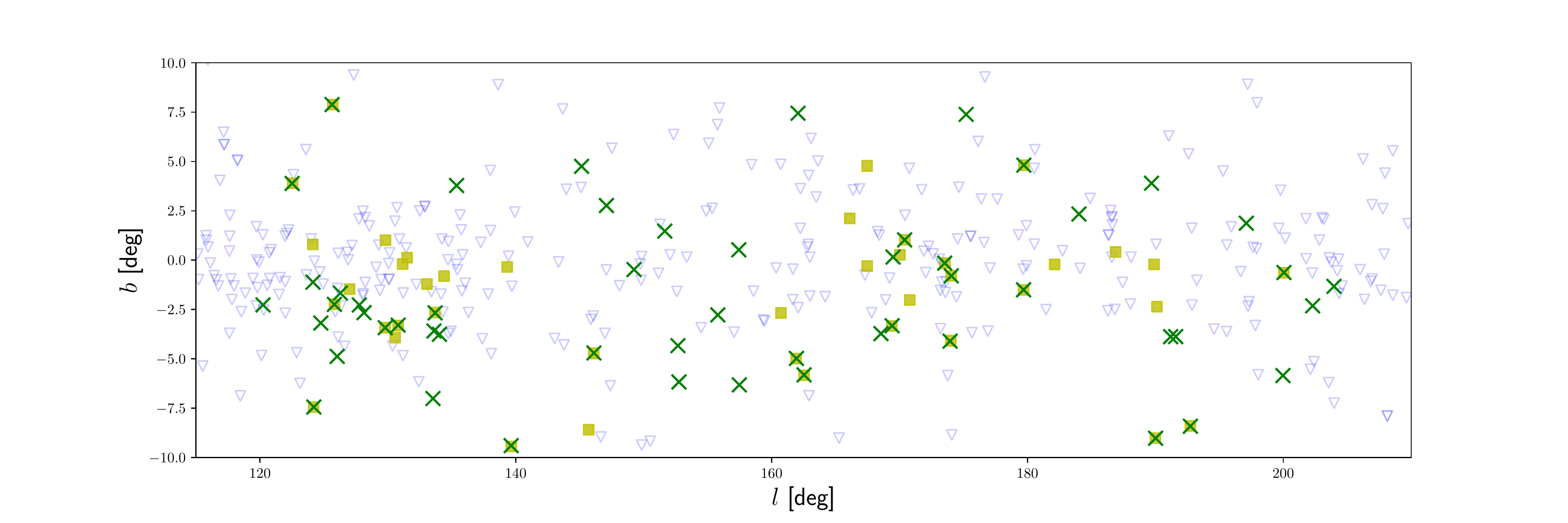}
\caption{Spatial distribution $(l,b)$ of the detected (green crosses) OCs, together with the already catalogued ones (blue triangles) in \citet{tristan_catalogue} and the COIN-\textit{Gaia} clusters (yellow boxes) \citep{coin_clusters}.}
\label{fig:spacedistribution}
\end{figure*}

\begin{figure}
\centering
\includegraphics[width = 1.00\columnwidth]{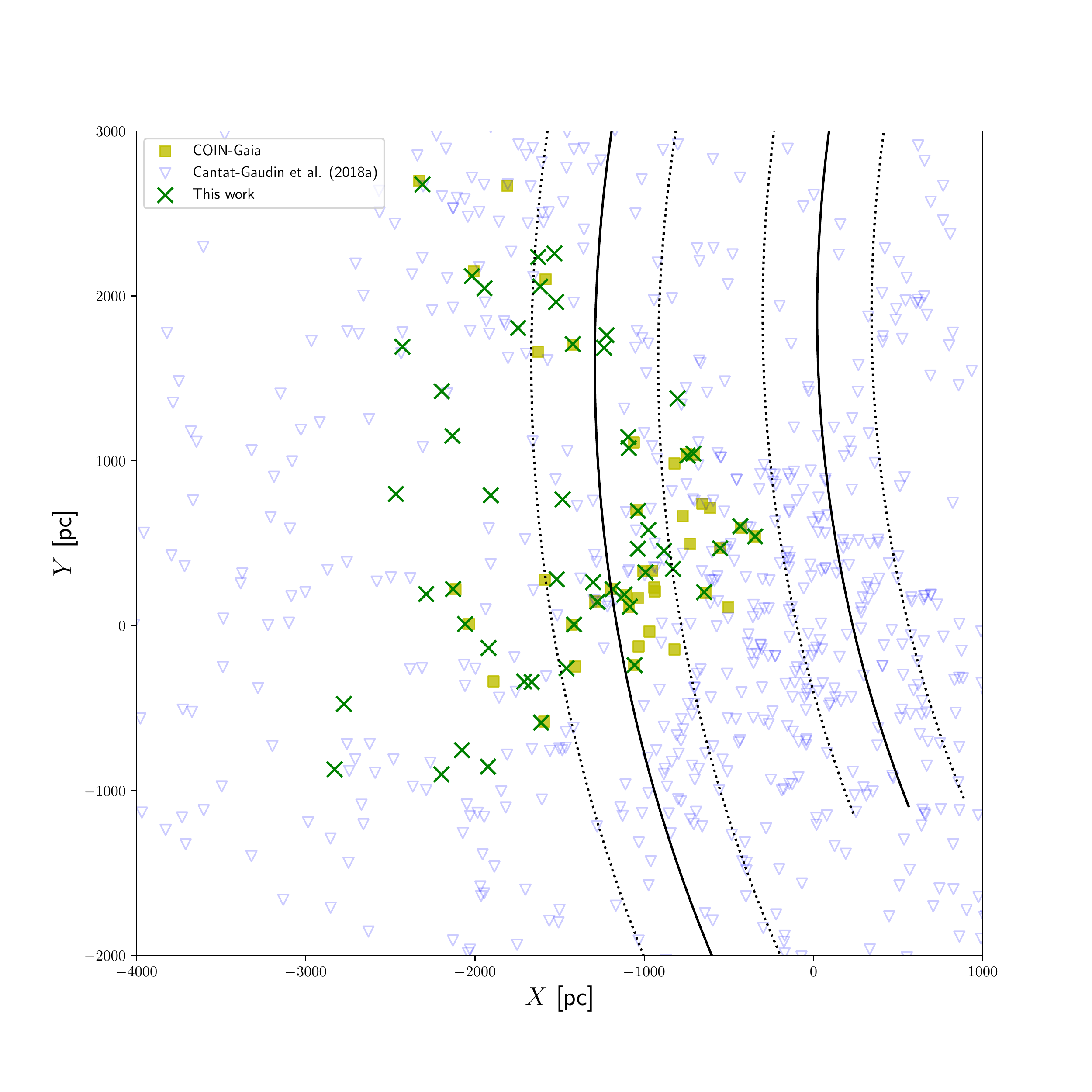}
\caption{X-Y projection of the detected OCs (green crosses)  together with already catalogued OCs (blue triangles) and COIN-\textit{Gaia} clusters (yellow boxes). Black lines represent the Local and the Perseus arms, plotted following the model in \citet{2014ApJ...783..130R}. The Sun is at $(0,0)$.}
\label{fig:xydensity}
\end{figure}

The newly found OCs are distributed along the Galactic anticentre direction as shown in \figref{fig:spacedistribution}, where green crosses represent OCs found in this work, blue triangles are the already catalogued OCs in \citet{tristan_catalogue} and yellow boxes are the OCs in \citet{coin_clusters}. It is worth noting that in a region around $l \sim \ang{140}$ the density of OCs decreases in terms of  catalogued clusters and of newly detected ones. This confirms the findings in \citet{coin_clusters} that this region seems to be devoid of OCs. The low OC density is better seen in \figref{fig:xydensity}, where an X-Y projection is shown with the Sun at $(0,0)$, and it seems to be pointing in the direction of the Perseus arm (Local and Perseus arms follow the model of \citealt{2014ApJ...783..130R}). This region of relatively low density was first reported as a lack of OB stars in the \textit{Gaia} DR2 data, and dubbed the Gulf of Camelopardalis\footnote{\label{gulf}https://www.cosmos.esa.int/web/gaia/iow\_20180614}.

\begin{figure}
\centering
\includegraphics[width = 1.00\columnwidth]{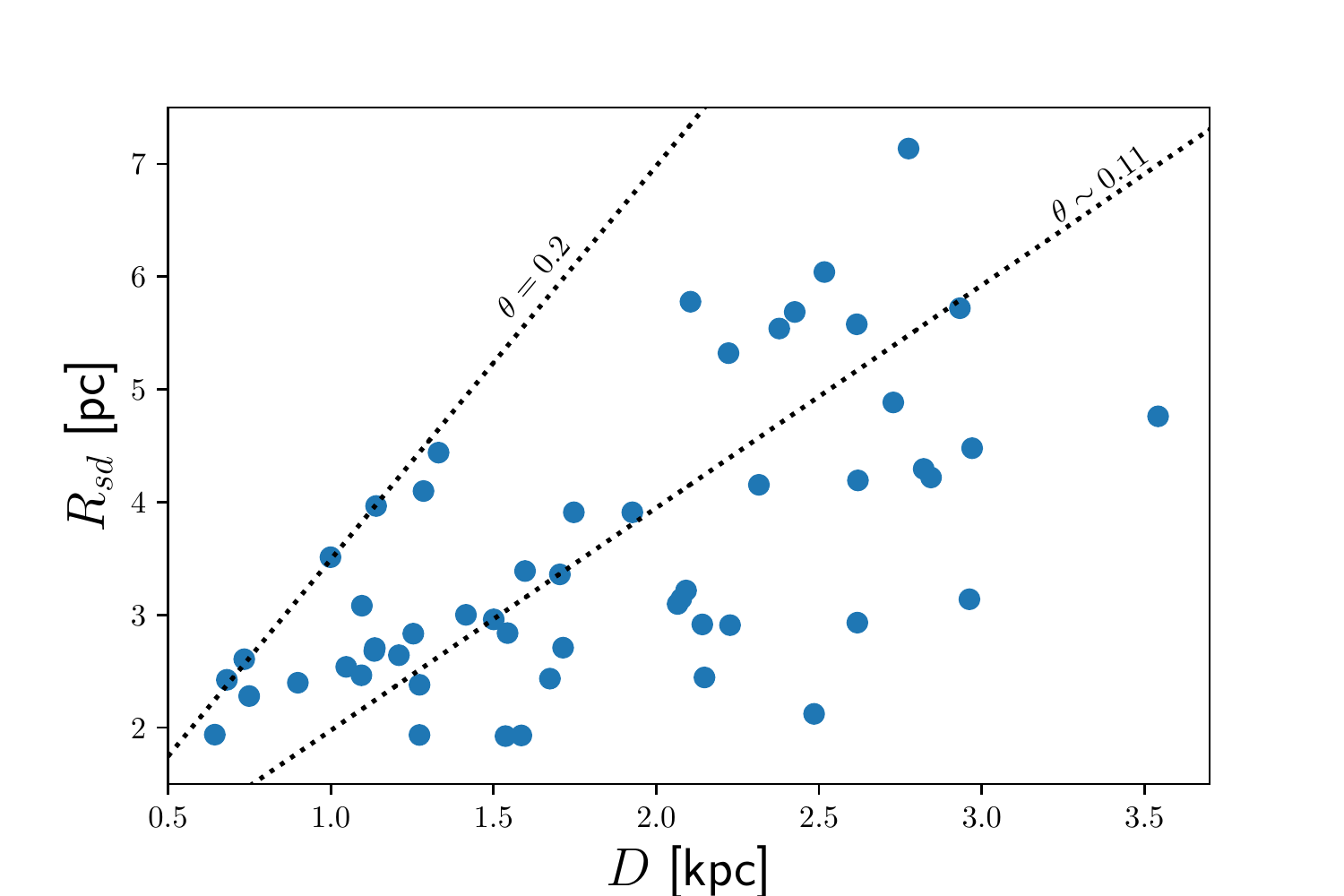}
\caption{Radius (computed from the standard deviation in $l$ and $b$ as $\sqrt{\sigma_l^2 + \sigma_b^2}$) as a function of distance for each of the reported $53$ OCs. Dotted lines represent the limiting cluster apparent size and the mean apparent size, \mbox{$\theta = \ang{0.2}$} and \mbox{$\theta \sim \ang{0.11}$}, respectively.}
\label{fig:RvsD}
\end{figure}

The  strategy we used to detect OCs relies on the OC field contrast, which is  able to detect those OCs with the highest contrast. This may result in a detection bias towards the more compact objects. \figrefalt{fig:RvsD} shows the radius of the detected OC as a function of its distance, which is computed as $1/\overline{\varpi}$ \citep{2018A&A...616A...9L} given the low parallax relative error ($\overline{\sigma_\varpi} \sim 0.04$ mas corresponding to $3-16\%$ in parallax relative error). The size range of the objects found  increases with distance, limiting our detection to very compact objects in a close neighbourhood. The mean size of the detected OCs is $\sigma_{l},\sigma_{b} \sim \ang{0.08}$, and  corresponds to an apparent size of $\theta \sim \ang{0.11}$. Our detection limit seems to be at a cluster apparent size of $\theta = \ang{0.2}$.

In terms of estimated distance, we find $6$ new OCs within $1$ kpc (the closest one at around $645$ pc) and $30$ within $1.8$ kpc, to be added to the $23$ found by \citet{acastro1} and the $31$ by \citet{coin_clusters}  in that distance range, further supporting the claim that more objects are yet to be discovered in this volume, especially with the combination of the excellent \textit{Gaia} data and ML algorithms in future all-sky searches. This challenges the statement that the OC census is complete up to $1.8$ kpc \citep{kharchenko,2018SSRv..214...74M,2018A&A...614A..22P}.

From the kinematical point of view, the reported OCs have a mean dispersion of $\mu_{\alpha^{*}},\mu_\delta \sim 0.2$ mas$\cdot$yr$^{-1}$, computed from the found member stars. This corresponds to a mean tangential velocity dispersion of $\sim 2.2$ km$\cdot$s$^{-1}$. Only $30$ of the $53$ reported OCs have a radial velocity measurement available in \textit{Gaia} DR2, $11$ of which have more than  two measurements $(N_{V_{rad}} > 2)$. The large $\sigma_{V_{rad}}$ for six of them may indicate the presence of binaries or non-members. Cross-matching with external surveys dedicated to radial velocity estimation, such as APOGEE \citep{2017AJ....154...94M}, does not add information (only one star was found in common between the two catalogues). The little information on radial velocities makes it difficult to characterise  OC members free of contamination from field stars.

\begin{figure*}
\begin{subfigure}{1.00\textwidth}
\centering
\includegraphics[scale = .35]{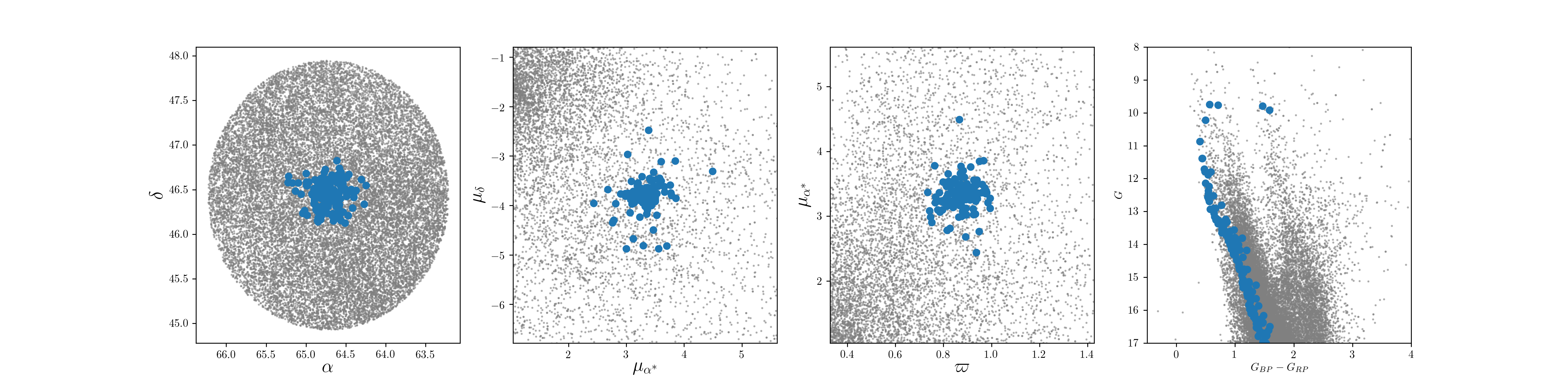}
\end{subfigure}
\begin{subfigure}{1.00\textwidth}
\centering
\includegraphics[scale = .35]{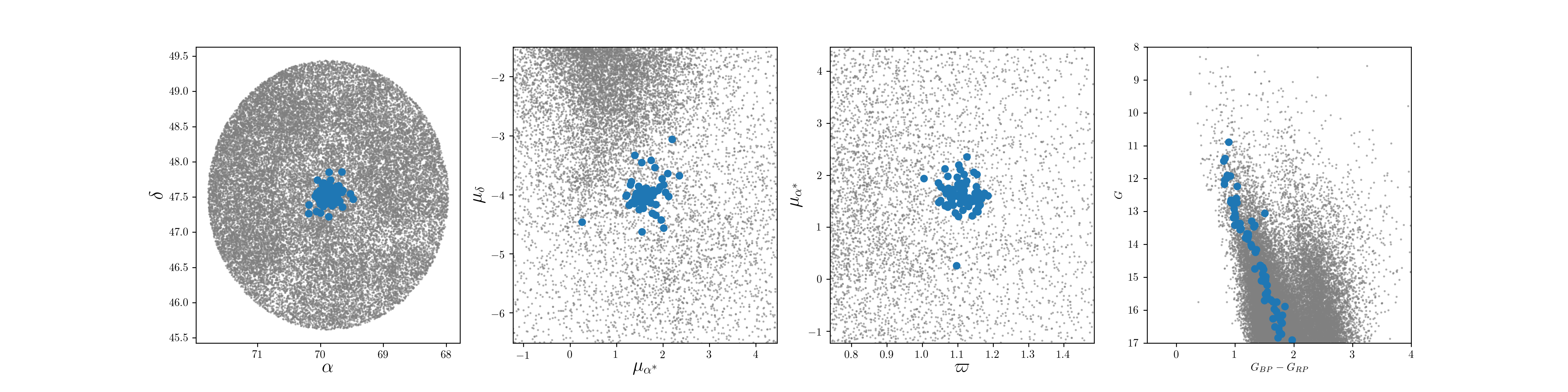}
\end{subfigure}
\begin{subfigure}{1.00\textwidth}
\centering
\includegraphics[scale = .35]{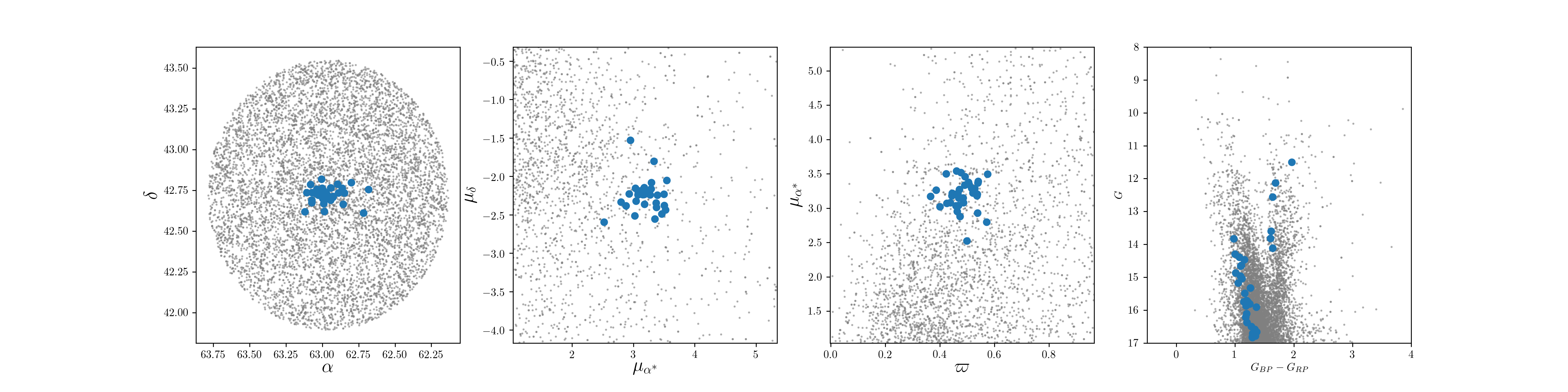}
\end{subfigure}
\begin{subfigure}{1.00\textwidth}
\centering
\includegraphics[scale = .35]{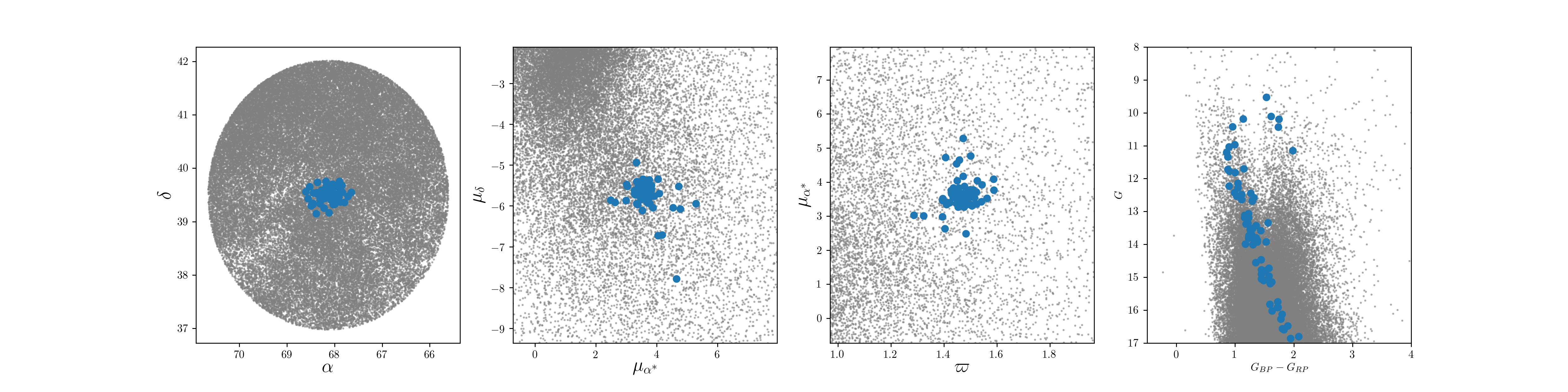}
\end{subfigure}
\begin{subfigure}{1.00\textwidth}
\centering
\includegraphics[scale = .35]{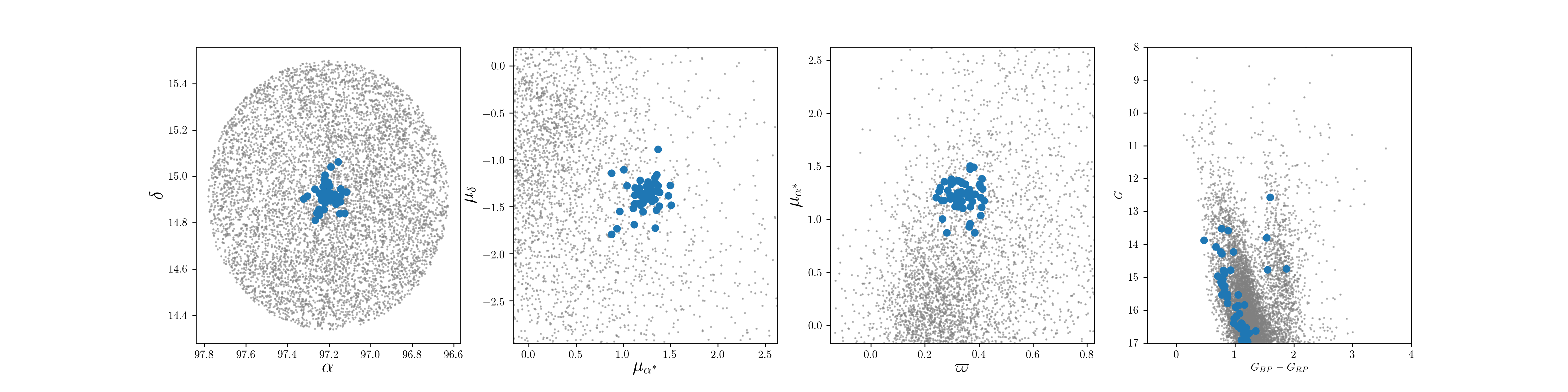}
\end{subfigure}
\caption{Five examples of the $53$ detected OCs. The blue dots represent the detected members, while grey dots represent field stars. The leftmost plots show the position of the OC in $(\alpha,\delta)$. The inner left plots show the $(\mu_{\alpha^*},\mu_\delta)$ distribution, whilst the inner right plots show the   $(\varpi,\mu_{\alpha^*})$ distribution. The rightmost plots show the CMD of each OC. The plotted OCs are, from top to bottom: \object{UBC~54}, \object{UBC~56}, \object{UBC~57}, \object{UBC~60}, and \object{UBC~90}. The first four clusters are class A and the last one is a class B cluster (see Table \ref{tab:new_oc}).}
\label{fig:CMD}
\end{figure*}

The photometric information is included when deciding if a CMD matches a real OC or not. This is done using an ANN trained with CMDs from the $1229$ OCs in \citet{tristan_catalogue} (see \secref{sec:method}), so the expected isochrone patterns are similar to those present in the training set. The ages of the reference clusters used in the training span from $40$ Myr to $1.5$ Gyr, so objects accepted by the ANN are in that age range. No estimation of photometric derived quantities is done here, only to mention that $25$ of the $53$ reported OCs have stars evolved beyond the main sequence, representing the oldest population of the found clusters. In \figref{fig:CMD} a few examples of detected OCs are shown,  four  class A and one class B, showing different ages. Together with the distribution in the five astrometric parameters $(\alpha,\delta,\varpi,\mu_{\alpha^*},\mu_\delta),$ the rightmost plots show the CMDs for each example OC.

\subsection{Matches with other catalogues}
\label{subsec:coincidences}

The candidates have been cross-matched with known catalogues of OCs \citep{dias,kharchenko}. These catalogues contain around $2000$ and $3000$ known stellar structures, respectively. However, some of these structures have  recently been found not to be real OCs \citep{2016AJ....152....7H,2018MNRAS.480.5242K,tristan_catalogue,2019A&A...624A...8A}. Moreover, the  catalogues were both  compiled from heterogeneous data sources, making the identification of an OC less reliable beyond positional arguments. We consider an OC to be positionally matched to a catalogued one if their centres lie within a circle of radius $r = \ang{0.5}$. We find $17$ candidates whose centres are identified with one object either from \citet{dias} or \citet{kharchenko}; however, none of the identifications are compatible in the rest of the astrometric mean parameters $(\varpi,\mu_{\alpha^*},\mu_{\delta})$, with the closest pair differing by $\sim 8\sigma$ in at least one parameter. However, we find \object{UBC~84} near the association Cas OB1, to which it may be related due to the extended region of  association. 

A recent list of $10\,978$ star clusters, associations, and candidates in the Milky Way has been published by \citet{2019AJ....157...12B}. Our list of candidates was cross-matched and only \object{UBC~90} and \object{UBC~44} are near one entry in the catalogue, Teutsch 20 and Patchick 12, respectively. Teutsch 20 and Patchick 12 are not listed in any of the other studied catalogues. Moreover, the quoted distance for Teutsch 20 is $2.54 \pm 0.05$ kpc \citep{2018RAA....18...32G} and we find \object{UBC~90} at $2.94$ kpc, which is not  compatible within errors. For the case of Patchick 12, we found no record of its mean astrometric parameters in the literature.

As said before, \citet{coin_clusters} recently found $41$ OCs located in roughly the same area of the sky, exploring the data of \textit{Gaia} DR2. We find $21$ OCs in common that share the five astrometric parameters (see \tabref{tab:new_oc}). The other $20$ OCs were not detected in our blind search, but we were able to recover them by increasing the OC field contrast when running DBSCAN in a cone search centred on the targeted OC. This shows that ML methods are complementary to each other, with none of the explored methods being able to detect  $100\%$ of the existing structures.

\section{Conclusions}
\label{sec:conclusions}

We use the methodology described in CG18 to systematically explore the \textit{Gaia} DR2 archive to search for unknown OCs in the anticentre direction. The method is a fully automated data mining task that uses an unsupervised clustering algorithm, DBSCAN, to find groups of stars that share common $(l,b,\varpi,\mu_{\alpha^*},\mu_{\delta})$ and decide whether or not they are real OCs based on an isochrone pattern recognition in the CMD using an ANN.

We can assess the overall performance in terms of the detection of already existing OCs. In this case, the method is able to find more than $75\%$ of the confirmed OCs in the search region. Most of the remaining $\sim 24\%$ of the clusters not found are recovered when the search is focused on the targeted OC. This suggests that our method works better for the OCs whose OC field contrast is high, and may be biased towards the more compact objects when the distance decreases.

The application of the whole methodology leads to the report of $53$ new OCs in a region covering the Galactic anticentre and the Perseus arm in the \textit{Gaia} DR2 data $(\ang{120} \leq l \leq \ang{205}$ and $\ang{-10} \leq b \leq \ang{10})$, which represents an increase of more than $22\%$ with respect to the OCs catalogued in this area. Moreover, $29$ of the detected OCs are closer than $2$ kpc, suggesting that there may be more groups to be detected in this volume. 

The density of OCs decreases in a region near $l \sim \ang{140}$. Very few OCs are found in this region, including already catalogued OCs and the newly reported OCs. This region has been named  the Gulf of Camelopardalis, and it reveals a complex structure of the second Galactic quadrant whose mapping was only recently made possible by \textit{Gaia} DR2 data, and still deserves further study.

The application of our methodology in the search regions shows that the census of OCs may not be complete. Moreover, other similar methodologies exploring the same region are able to find more groups not detected via our method, while they missed some groups detected here. We conclude that a blind search using a single detection method is not able not recover all the existing stellar structures, and that different ML algorithms for this purpose are complementary to each other.

The design of the whole methodology, requiring minimal manual intervention, means that  its application to a big data set such as the whole \textit{Gaia} DR2 is possible. The planned future exploitation of the \textit{Gaia} archive in terms of blind search of OCs would represent a huge increase to the known OC population.

\begin{acknowledgements}

This work has made use of results from the European Space Agency (ESA)
space mission {\it Gaia}, the data from which were processed by the {\it Gaia
Data Processing and Analysis Consortium} (DPAC).  Funding for the DPAC
has been provided by national institutions, in particular the
institutions participating in the {\it Gaia} Multilateral Agreement. The
{\it Gaia} mission website is \url{http: //www.cosmos.esa.int/gaia}. The
authors are current or past members of the ESA {\it Gaia} mission team and
of the {\it Gaia} DPAC.

This work was supported by the MINECO (Spanish Ministry of Economy) through 
grant ESP2016-80079-C2-1-R (MINECO/FEDER, UE) and MDM-2014-0369 of ICCUB (Unidad de Excelencia ``María de Maeztu''). 

This research has made use of the TOPCAT \citep{topcat}.
This research has made use of the VizieR catalogue access tool, CDS,
Strasbourg, France. The original description of the VizieR service was
published in A$\&$AS 143, 23.

ACG thanks Dr. Laia Casamiquela for her useful comments.
\end{acknowledgements}

\bibliographystyle{aa} 
\bibliography{bibliography} 

\begin{thebibliography}{34}
\expandafter\ifx\csname natexlab\endcsname\relax\def\natexlab#1{#1}\fi

\bibitem[{{Angelo} {et~al.}(2019){Angelo}, {Santos}, {Corradi}, \&
  {Maia}}]{2019A&A...624A...8A}
{Angelo}, M.~S., {Santos}, J.~F.~C., {Corradi}, W.~J.~B., \& {Maia}, F.~F.~S.
  2019, \aap, 624, A8

\bibitem[{{Bica} {et~al.}(2019){Bica}, {Pavani}, {Bonatto}, \&
  {Lima}}]{2019AJ....157...12B}
{Bica}, E., {Pavani}, D.~B., {Bonatto}, C.~J., \& {Lima}, E.~F. 2019, \aj, 157,
  12

\bibitem[{{Bossini} {et~al.}(2019){Bossini}, {Vallenari}, {Bragaglia},
  {Cantat-Gaudin}, {Sordo}, {Balaguer-N{\'u}{\~n}ez}, {Jordi}, {Moitinho},
  {Soubiran}, {Casamiquela}, {Carrera}, \& {Heiter}}]{2019A&A...623A.108B}
{Bossini}, D., {Vallenari}, A., {Bragaglia}, A., {et~al.} 2019, \aap, 623, A108

\bibitem[{{Cantat-Gaudin} {et~al.}(2018){Cantat-Gaudin}, {Jordi}, {Vallenari},
  {Bragaglia}, {Balaguer-N{\'u}{\~n}ez}, {Soubiran}, {Bossini}, {Moitinho},
  {Castro-Ginard}, {Krone-Martins}, {Casamiquela}, {Sordo}, \&
  {Carrera}}]{tristan_catalogue}
{Cantat-Gaudin}, T., {Jordi}, C., {Vallenari}, A., {et~al.} 2018, \aap, 618,
  A93

\bibitem[{{Cantat-Gaudin} {et~al.}(2019){Cantat-Gaudin}, {Krone-Martins},
  {Sedaghat}, {Farahi}, {de Souza}, {Skalidis}, {Malz}, {Mac{\^e}do}, {Moews},
  {Jordi}, {Moitinho}, {Castro-Ginard}, {Ishida}, {Heneka}, {Boucaud}, \&
  {Trindade}}]{coin_clusters}
{Cantat-Gaudin}, T., {Krone-Martins}, A., {Sedaghat}, N., {et~al.} 2019, \aap,
  624, A126

\bibitem[{{Castro-Ginard} {et~al.}(2018){Castro-Ginard}, {Jordi}, {Luri},
  {Julbe}, {Morvan}, {Balaguer-N{\'u}{\~n}ez}, \& {Cantat- Gaudin}}]{acastro1}
{Castro-Ginard}, A., {Jordi}, C., {Luri}, X., {et~al.} 2018, \aap, 618, A59

\bibitem[{{Dias} {et~al.}(2002){Dias}, {Alessi}, {Moitinho}, \&
  {L{\'e}pine}}]{dias}
{Dias}, W.~S., {Alessi}, B.~S., {Moitinho}, A., \& {L{\'e}pine}, J.~R.~D. 2002,
  \aap, 389, 871

\bibitem[{Ester {et~al.}(1996)Ester, Kriegel, Sander, \& Xu}]{dbscan}
Ester, M., Kriegel, H.-P., Sander, J., \& Xu, X. 1996, in Proceedings of the
  Second International Conference on Knowledge Discovery and Data Mining,
  KDD'96 (AAAI Press), 226--231

\bibitem[{{Evans} {et~al.}(2018){Evans}, {Riello}, {De Angeli}, {Carrasco},
  {Montegriffo}, {Fabricius}, {Jordi}, {Palaversa}, {Diener}, {Busso},
  {Cacciari}, {van Leeuwen}, {Burgess}, {Davidson}, {Harrison}, {Hodgkin},
  {Pancino}, {Richards}, {Altavilla}, {Balaguer-N{\'u}{\~n}ez}, {Barstow},
  {Bellazzini}, {Brown}, {Castellani}, {Cocozza}, {De Luise}, {Delgado},
  {Ducourant}, {Galleti}, {Gilmore}, {Giuffrida}, {Holl}, {Kewley}, {Koposov},
  {Marinoni}, {Marrese}, {Osborne}, {Piersimoni}, {Portell}, {Pulone},
  {Ragaini}, {Sanna}, {Terrett}, {Walton}, {Wevers}, \&
  {Wyrzykowski}}]{2018A&A...616A...4E}
{Evans}, D.~W., {Riello}, M., {De Angeli}, F., {et~al.} 2018, \aap, 616, A4

\bibitem[{{Froebrich} {et~al.}(2007){Froebrich}, {Scholz}, \&
  {Raftery}}]{2007MNRAS.374..399F}
{Froebrich}, D., {Scholz}, A., \& {Raftery}, C.~L. 2007, \mnras, 374, 399

\bibitem[{{Gaia Collaboration} {et~al.}(2018){Gaia Collaboration}, {Brown},
  {Vallenari}, {Prusti}, {de Bruijne}, {Babusiaux}, {Bailer-Jones}, {Biermann},
  {Evans}, {Eyer}, \& et~al.}]{2018A&A...616A...1G}
{Gaia Collaboration}, {Brown}, A.~G.~A., {Vallenari}, A., {et~al.} 2018, \aap,
  616, A1

\bibitem[{{Gaia Collaboration} {et~al.}(2016){Gaia Collaboration}, {Prusti},
  {de Bruijne}, {Brown}, {Vallenari}, {Babusiaux}, {Bailer-Jones}, {Bastian},
  {Biermann}, {Evans}, \& et~al.}]{2016A&A...595A...1G}
{Gaia Collaboration}, {Prusti}, T., {de Bruijne}, J.~H.~J., {et~al.} 2016,
  \aap, 595, A1

\bibitem[{{Gao}(2018{\natexlab{a}})}]{2018ApJ...869....9G}
{Gao}, X. 2018{\natexlab{a}}, \apj, 869, 9

\bibitem[{{Gao}(2018{\natexlab{b}})}]{2018Ap&SS.363..232G}
{Gao}, X.-H. 2018{\natexlab{b}}, \apss, 363, 232

\bibitem[{{Guo} {et~al.}(2018){Guo}, {Zhang}, {Zhang}, {Liu}, {Yuan}, {Huang},
  {Wang}, {Chen}, {Zhao}, {Liu}, {Chen}, {Xiang}, {Tian}, {Huo}, \&
  {Wang}}]{2018RAA....18...32G}
{Guo}, J.-C., {Zhang}, H.-W., {Zhang}, H.-H., {et~al.} 2018, Research in
  Astronomy and Astrophysics, 18, 032

\bibitem[{{Han} {et~al.}(2016){Han}, {Curtis}, \&
  {Wright}}]{2016AJ....152....7H}
{Han}, E., {Curtis}, J.~L., \& {Wright}, J.~T. 2016, \aj, 152, 7

\bibitem[{Hinton(1989)}]{ann}
Hinton, G. 1989, Artificial Intelligence, 40, 185

\bibitem[{{Kharchenko} {et~al.}(2013){Kharchenko}, {Piskunov}, {Schilbach},
  {R{\"o}ser}, \& {Scholz}}]{kharchenko}
{Kharchenko}, N.~V., {Piskunov}, A.~E., {Schilbach}, E., {R{\"o}ser}, S., \&
  {Scholz}, R.-D. 2013, \aap, 558, A53

\bibitem[{{Kos} {et~al.}(2018){Kos}, {de Silva}, {Buder}, {Bland -Hawthorn},
  {Sharma}, {Asplund}, {D'Orazi}, {Duong}, {Freeman}, {Lewis}, {Lin}, {Lind},
  {Martell}, {Schlesinger}, {Simpson}, {Zucker}, {Zwitter}, {Bedding},
  {{\v{C}}otar}, {Horner}, {Nordlander}, {Stello}, {Ting}, \&
  {Traven}}]{2018MNRAS.480.5242K}
{Kos}, J., {de Silva}, G., {Buder}, S., {et~al.} 2018, \mnras, 480, 5242

\bibitem[{{Krone-Martins} \& {Moitinho}(2014)}]{2014A&A...561A..57K}
{Krone-Martins}, A. \& {Moitinho}, A. 2014, \aap, 561, A57

\bibitem[{{Lindegren} {et~al.}(2018){Lindegren}, {Hern{\'a}ndez}, {Bombrun},
  {Klioner}, {Bastian}, {Ramos-Lerate}, {de Torres}, {Steidelm{\"u}ller},
  {Stephenson}, {Hobbs}, {Lammers}, {Biermann}, {Geyer}, {Hilger}, {Michalik},
  {Stampa}, {McMillan}, {Casta{\~n}eda}, {Clotet}, {Comoretto}, {Davidson},
  {Fabricius}, {Gracia}, {Hambly}, {Hutton}, {Mora}, {Portell}, {van Leeuwen},
  {Abbas}, {Abreu}, {Altmann}, {Andrei}, {Anglada}, {Balaguer-N{\'u}{\~n}ez},
  {Barache}, {Becciani}, {Bertone}, {Bianchi}, {Bouquillon}, {Bourda},
  {Br{\"u}semeister}, {Bucciarelli}, {Busonero}, {Buzzi}, {Cancelliere},
  {Carlucci}, {Charlot}, {Cheek}, {Crosta}, {Crowley}, {de Bruijne}, {de
  Felice}, {Drimmel}, {Esquej}, {Fienga}, {Fraile}, {Gai}, {Garralda},
  {Gonz{\'a}lez-Vidal}, {Guerra}, {Hauser}, {Hofmann}, {Holl}, {Jordan},
  {Lattanzi}, {Lenhardt}, {Liao}, {Licata}, {Lister}, {L{\"o}ffler},
  {Marchant}, {Martin-Fleitas}, {Messineo}, {Mignard}, {Morbidelli}, {Poggio},
  {Riva}, {Rowell}, {Salguero}, {Sarasso}, {Sciacca}, {Siddiqui}, {Smart},
  {Spagna}, {Steele}, {Taris}, {Torra}, {van Elteren}, {van Reeven}, \&
  {Vecchiato}}]{2018A&A...616A...2L}
{Lindegren}, L., {Hern{\'a}ndez}, J., {Bombrun}, A., {et~al.} 2018, \aap, 616,
  A2

\bibitem[{{Lindegren} {et~al.}(2016){Lindegren}, {Lammers}, {Bastian},
  {Hern{\'a}ndez}, {Klioner}, {Hobbs}, {Bombrun}, {Michalik}, {Ramos-Lerate},
  {Butkevich}, {Comoretto}, {Joliet}, {Holl}, {Hutton}, {Parsons},
  {Steidelm{\"u}ller}, {Abbas}, {Altmann}, {Andrei}, {Anton}, {Bach},
  {Barache}, {Becciani}, {Berthier}, {Bianchi}, {Biermann}, {Bouquillon},
  {Bourda}, {Br{\"u}semeister}, {Bucciarelli}, {Busonero}, {Carlucci},
  {Casta{\~n}eda}, {Charlot}, {Clotet}, {Crosta}, {Davidson}, {de Felice},
  {Drimmel}, {Fabricius}, {Fienga}, {Figueras}, {Fraile}, {Gai}, {Garralda},
  {Geyer}, {Gonz{\'a}lez-Vidal}, {Guerra}, {Hambly}, {Hauser}, {Jordan},
  {Lattanzi}, {Lenhardt}, {Liao}, {L{\"o}ffler}, {McMillan}, {Mignard}, {Mora},
  {Morbidelli}, {Portell}, {Riva}, {Sarasso}, {Serraller}, {Siddiqui}, {Smart},
  {Spagna}, {Stampa}, {Steele}, {Taris}, {Torra}, {van Reeven}, {Vecchiato},
  {Zschocke}, {de Bruijne}, {Gracia}, {Raison}, {Lister}, {Marchant},
  {Messineo}, {Soffel}, {Osorio}, {de Torres}, \& {O'Mullane}}]{gdr1-tgas}
{Lindegren}, L., {Lammers}, U., {Bastian}, U., {et~al.} 2016, \aap, 595, A4

\bibitem[{{Luri} {et~al.}(2018){Luri}, {Brown}, {Sarro}, {Arenou},
  {Bailer-Jones}, {Castro-Ginard}, {de Bruijne}, {Prusti}, {Babusiaux}, \&
  {Delgado}}]{2018A&A...616A...9L}
{Luri}, X., {Brown}, A.~G.~A., {Sarro}, L.~M., {et~al.} 2018, \aap, 616, A9

\bibitem[{{Majewski} {et~al.}(2017){Majewski}, {Schiavon}, {Frinchaboy},
  {Allende Prieto}, {Barkhouser}, {Bizyaev}, {Blank}, {Brunner}, {Burton},
  {Carrera}, {Chojnowski}, {Cunha}, {Epstein}, {Fitzgerald}, {Garc{\'\i}a
  P{\'e}rez}, {Hearty}, {Henderson}, {Holtzman}, {Johnson}, {Lam}, {Lawler},
  {Maseman}, {M{\'e}sz{\'a}ros}, {Nelson}, {Nguyen}, {Nidever}, {Pinsonneault},
  {Shetrone}, {Smee}, {Smith}, {Stolberg}, {Skrutskie}, {Walker}, {Wilson},
  {Zasowski}, {Anders}, {Basu}, {Beland}, {Blanton}, {Bovy}, {Brownstein},
  {Carlberg}, {Chaplin}, {Chiappini}, {Eisenstein}, {Elsworth}, {Feuillet},
  {Fleming}, {Galbraith-Frew}, {Garc{\'\i}a}, {Garc{\'\i}a-Hern{\'a}ndez},
  {Gillespie}, {Girardi}, {Gunn}, {Hasselquist}, {Hayden}, {Hekker}, {Ivans},
  {Kinemuchi}, {Klaene}, {Mahadevan}, {Mathur}, {Mosser}, {Muna}, {Munn},
  {Nichol}, {O'Connell}, {Parejko}, {Robin}, {Rocha-Pinto}, {Schultheis},
  {Serenelli}, {Shane}, {Silva Aguirre}, {Sobeck}, {Thompson}, {Troup},
  {Weinberg}, \& {Zamora}}]{2017AJ....154...94M}
{Majewski}, S.~R., {Schiavon}, R.~P., {Frinchaboy}, P.~M., {et~al.} 2017, \aj,
  154, 94

\bibitem[{{Matsunaga} {et~al.}(2018){Matsunaga}, {Bono}, {Chen}, {de Grijs},
  {Inno}, \& {Nishiyama}}]{2018SSRv..214...74M}
{Matsunaga}, N., {Bono}, G., {Chen}, X., {et~al.} 2018, \ssr, 214, 74

\bibitem[{{Michalik} {et~al.}(2015){Michalik}, {Lindegren}, \& {Hobbs}}]{tgas}
{Michalik}, D., {Lindegren}, L., \& {Hobbs}, D. 2015, \aap, 574, A115

\bibitem[{Pedregosa {et~al.}(2011)Pedregosa, Varoquaux, Gramfort, Michel,
  Thirion, Grisel, Blondel, Prettenhofer, Weiss, Dubourg, Vanderplas, Passos,
  Cournapeau, Brucher, Perrot, \& Duchesnay}]{sklearn}
Pedregosa, F., Varoquaux, G., Gramfort, A., {et~al.} 2011, Journal of Machine
  Learning Research, 12, 2825

\bibitem[{{Piskunov} {et~al.}(2018){Piskunov}, {Just}, {Kharchenko}, {Berczik},
  {Scholz}, {Reffert}, \& {Yen}}]{2018A&A...614A..22P}
{Piskunov}, A.~E., {Just}, A., {Kharchenko}, N.~V., {et~al.} 2018, \aap, 614,
  A22

\bibitem[{{Reid} {et~al.}(2014){Reid}, {Menten}, {Brunthaler}, {Zheng}, {Dame},
  {Xu}, {Wu}, {Zhang}, {Sanna}, {Sato}, {Hachisuka}, {Choi}, {Immer},
  {Moscadelli}, {Rygl}, \& {Bartkiewicz}}]{2014ApJ...783..130R}
{Reid}, M.~J., {Menten}, K.~M., {Brunthaler}, A., {et~al.} 2014, \apj, 783, 130

\bibitem[{{R{\"o}ser} {et~al.}(2016){R{\"o}ser}, {Schilbach}, \&
  {Goldman}}]{2016A&A...595A..22R}
{R{\"o}ser}, S., {Schilbach}, E., \& {Goldman}, B. 2016, \aap, 595, A22

\bibitem[{{Schmeja} {et~al.}(2014){Schmeja}, {Kharchenko}, {Piskunov},
  {R{\"o}ser}, {Schilbach}, {Froebrich}, \& {Scholz}}]{2014A&A...568A..51S}
{Schmeja}, S., {Kharchenko}, N.~V., {Piskunov}, A.~E., {et~al.} 2014, \aap,
  568, A51

\bibitem[{{Scholz} {et~al.}(2015){Scholz}, {Kharchenko}, {Piskunov},
  {R{\"o}ser}, \& {Schilbach}}]{2015A&A...581A..39S}
{Scholz}, R.~D., {Kharchenko}, N.~V., {Piskunov}, A.~E., {R{\"o}ser}, S., \&
  {Schilbach}, E. 2015, \aap, 581, A39

\bibitem[{{Soubiran} {et~al.}(2018){Soubiran}, {Cantat-Gaudin},
  {Romero-G{\'o}mez}, {Casamiquela}, {Jordi}, {Vallenari}, {Antoja},
  {Balaguer-N{\'u}{\~n}ez}, {Bossini}, {Bragaglia}, {Carrera}, {Castro-Ginard},
  {Figueras}, {Heiter}, {Katz}, {Krone- Martins}, {Le Campion}, {Moitinho}, \&
  {Sordo}}]{2018A&A...619A.155S}
{Soubiran}, C., {Cantat-Gaudin}, T., {Romero-G{\'o}mez}, M., {et~al.} 2018,
  \aap, 619, A155

\bibitem[{{Taylor}(2005)}]{topcat}
{Taylor}, M.~B. 2005, in Astronomical Society of the Pacific Conference Series,
  Vol. 347, Astronomical Data Analysis Software and Systems XIV, ed.
  P.~{Shopbell}, M.~{Britton}, \& R.~{Ebert}, 29

\end{thebibliography}

\end{document}